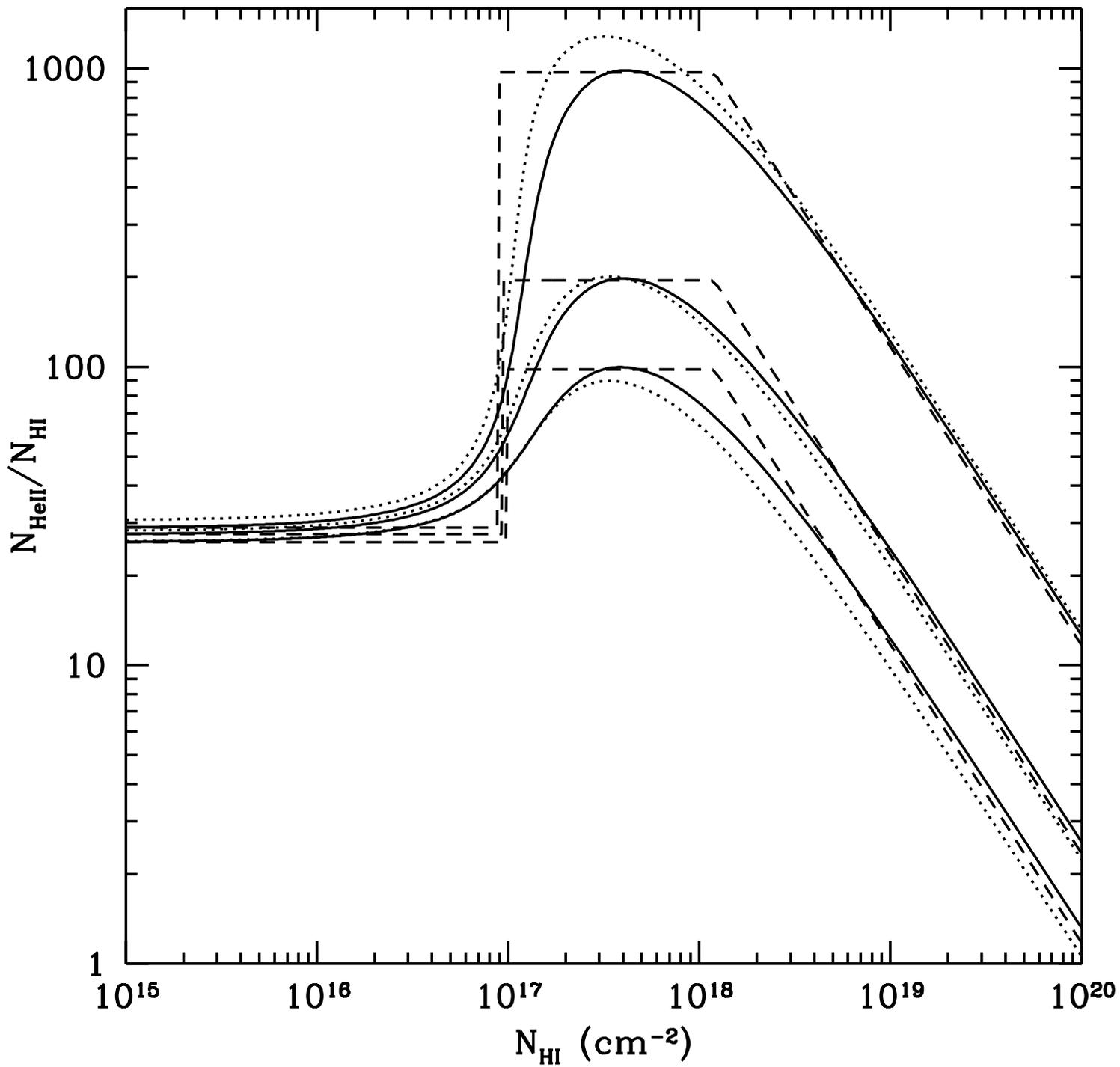

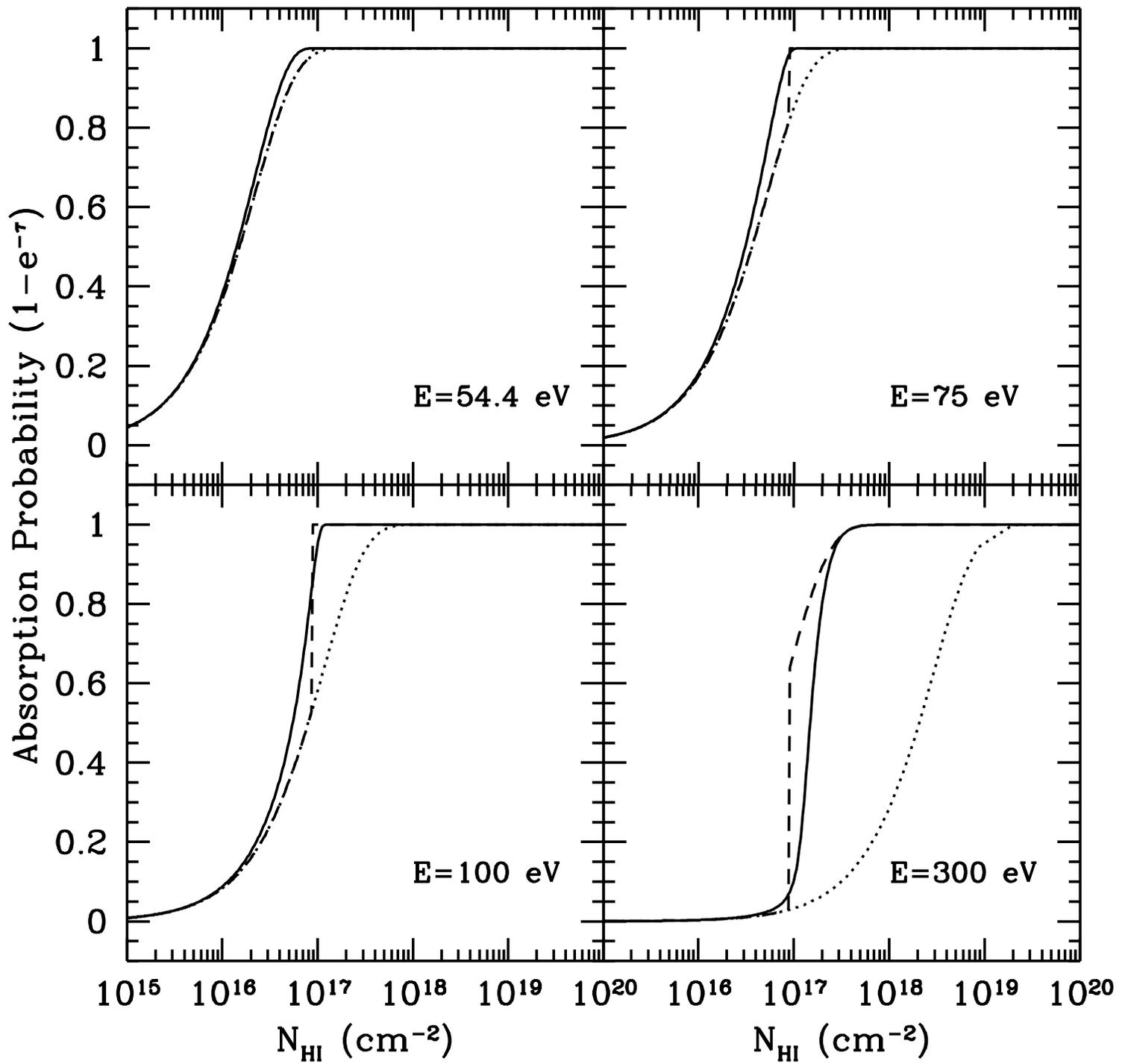

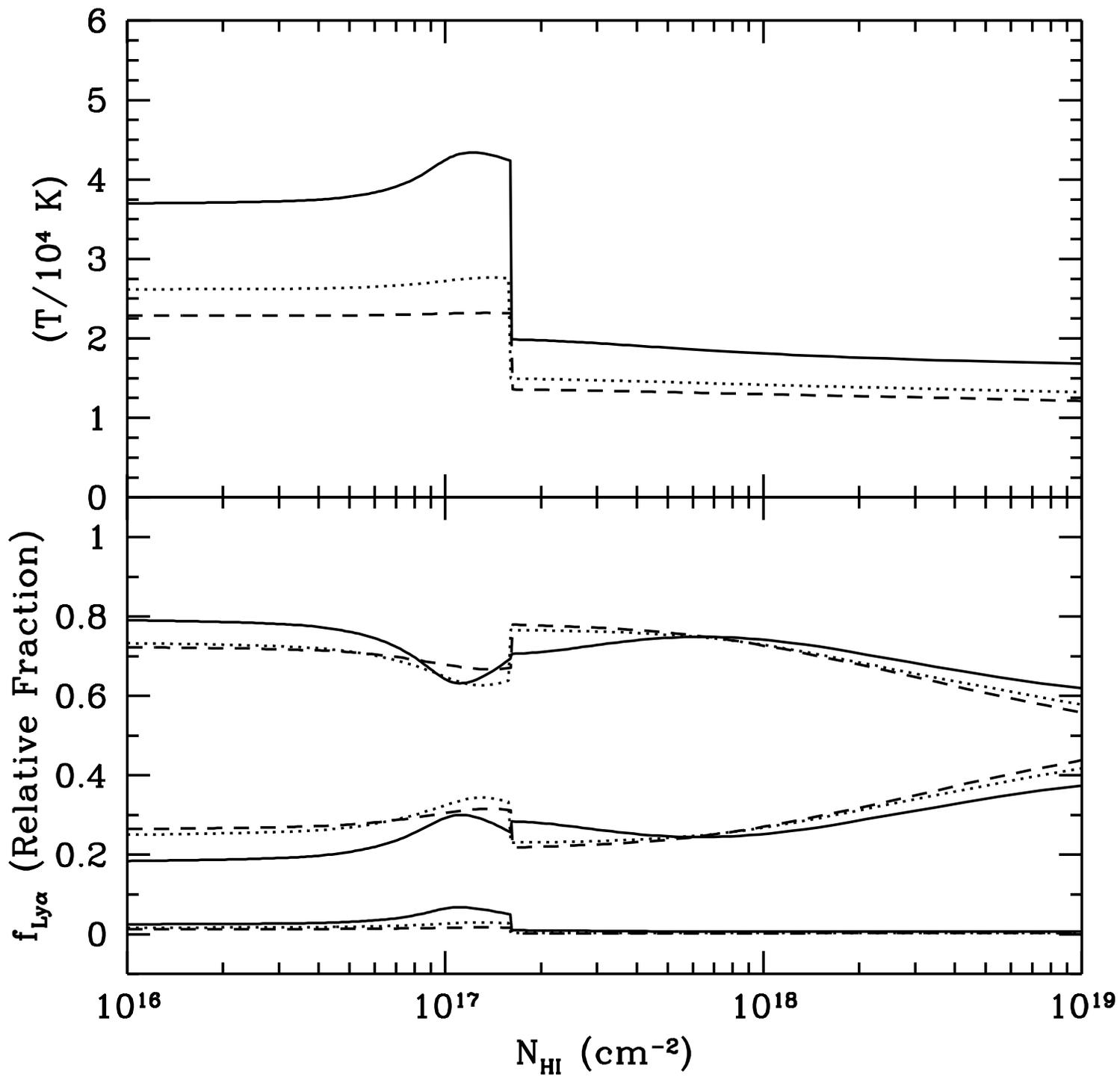

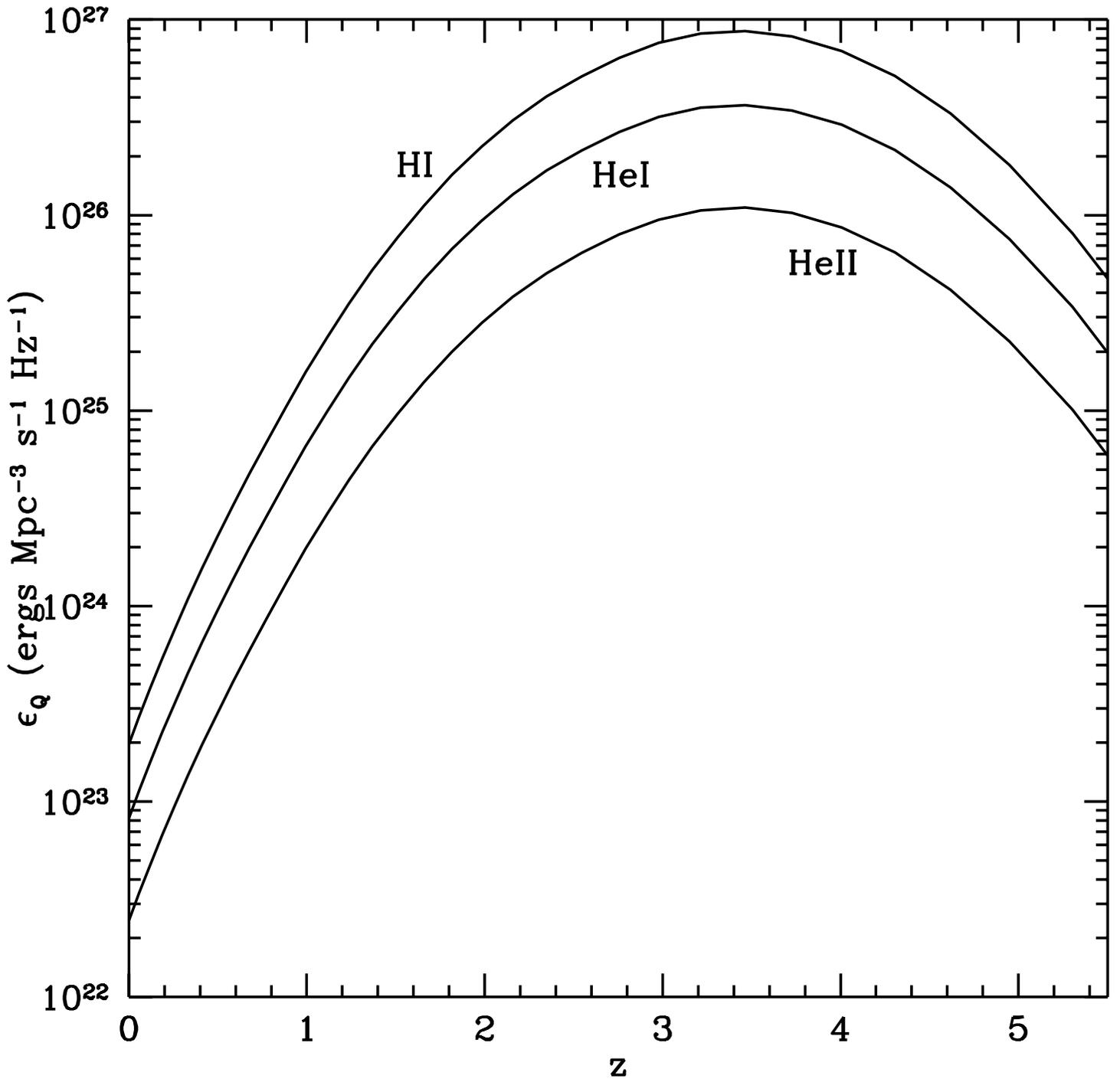

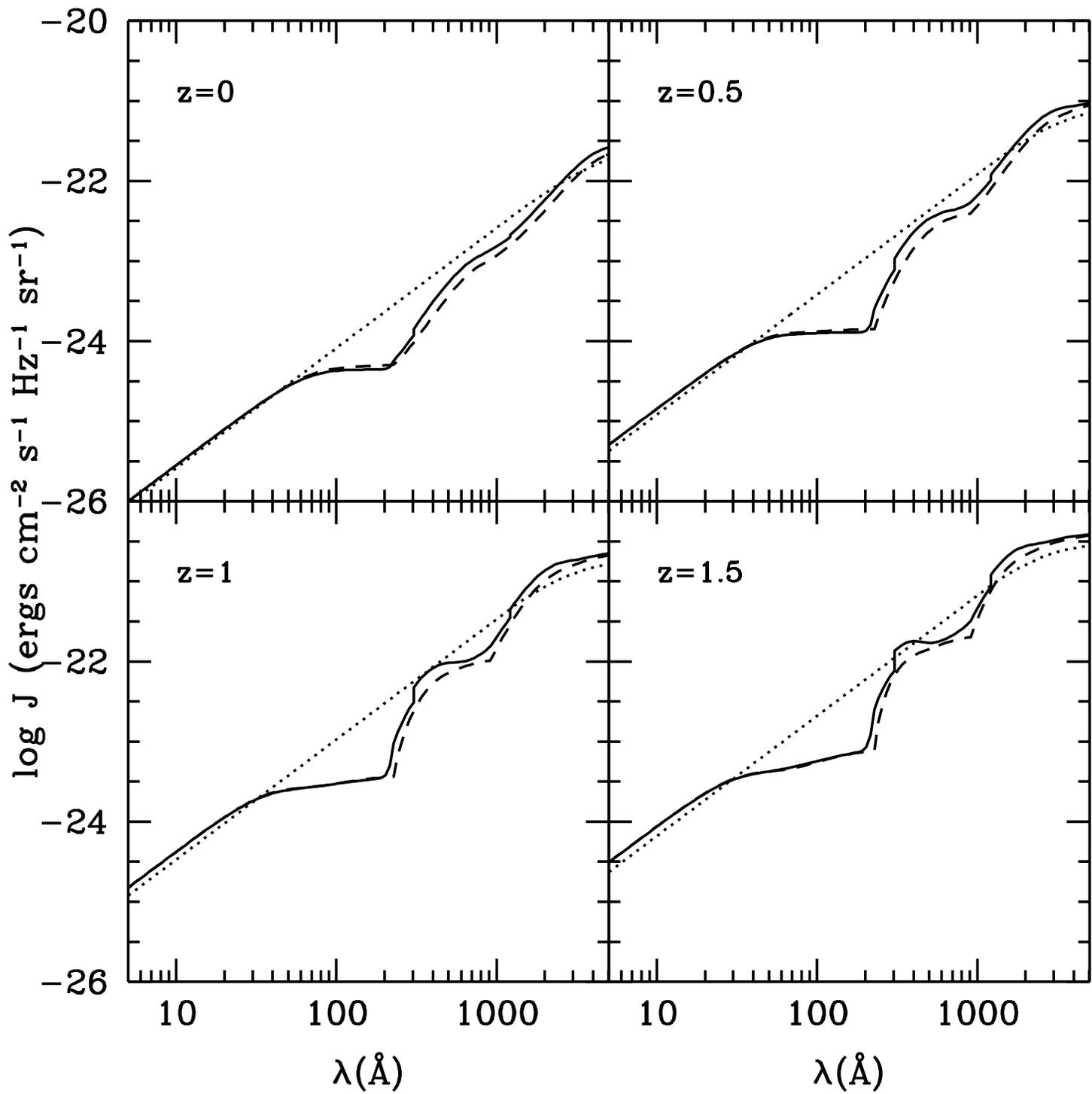

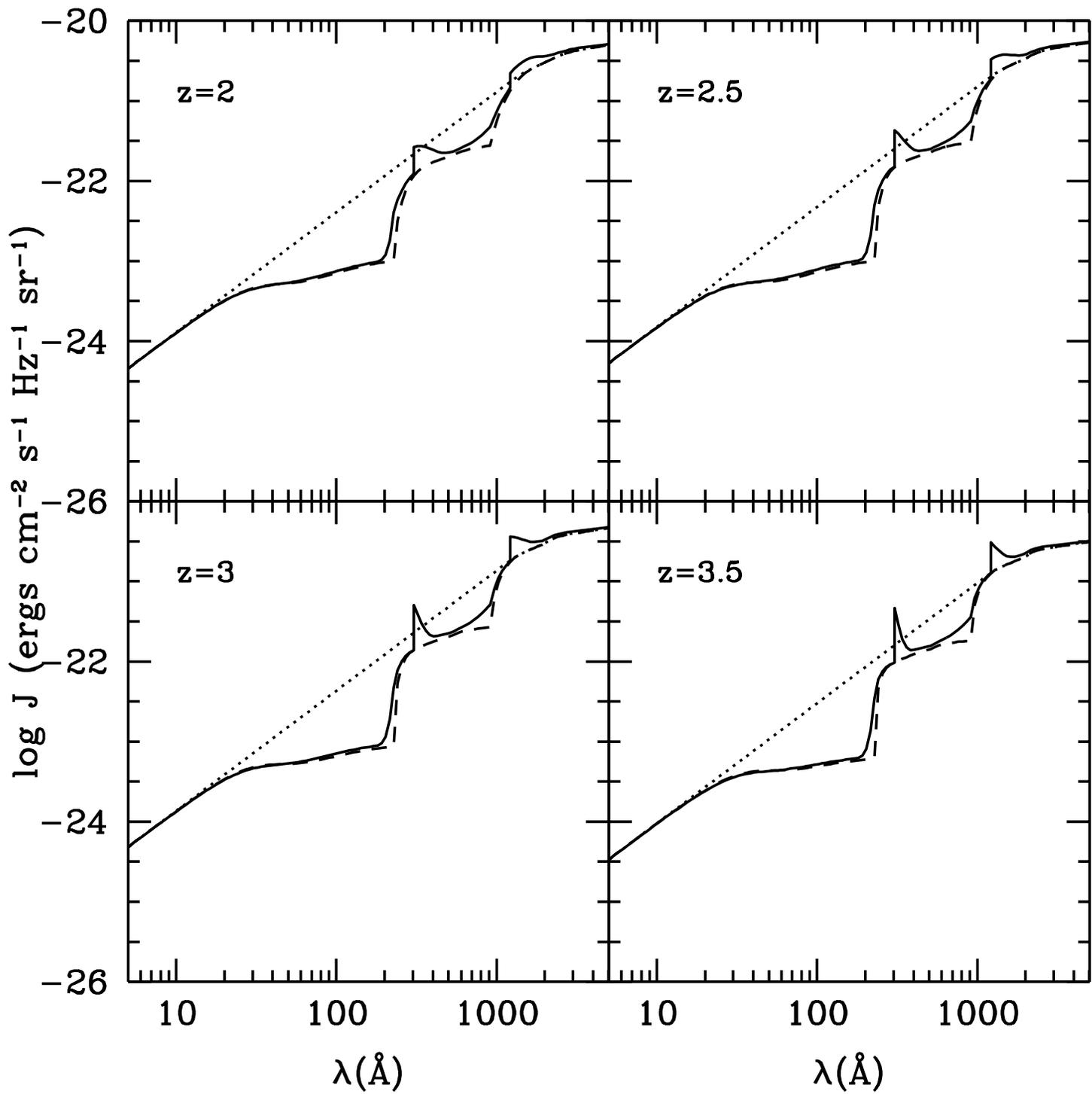

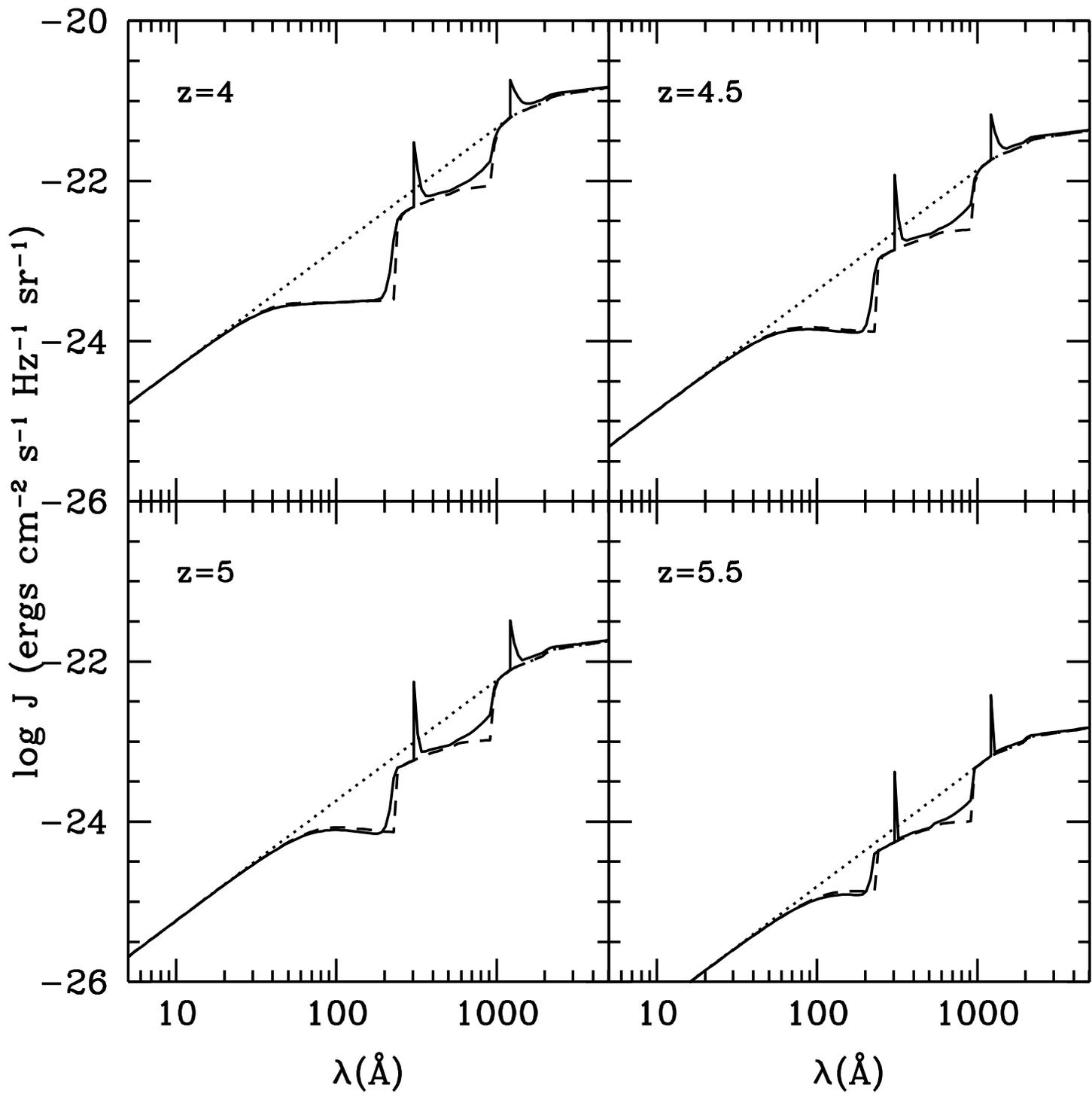

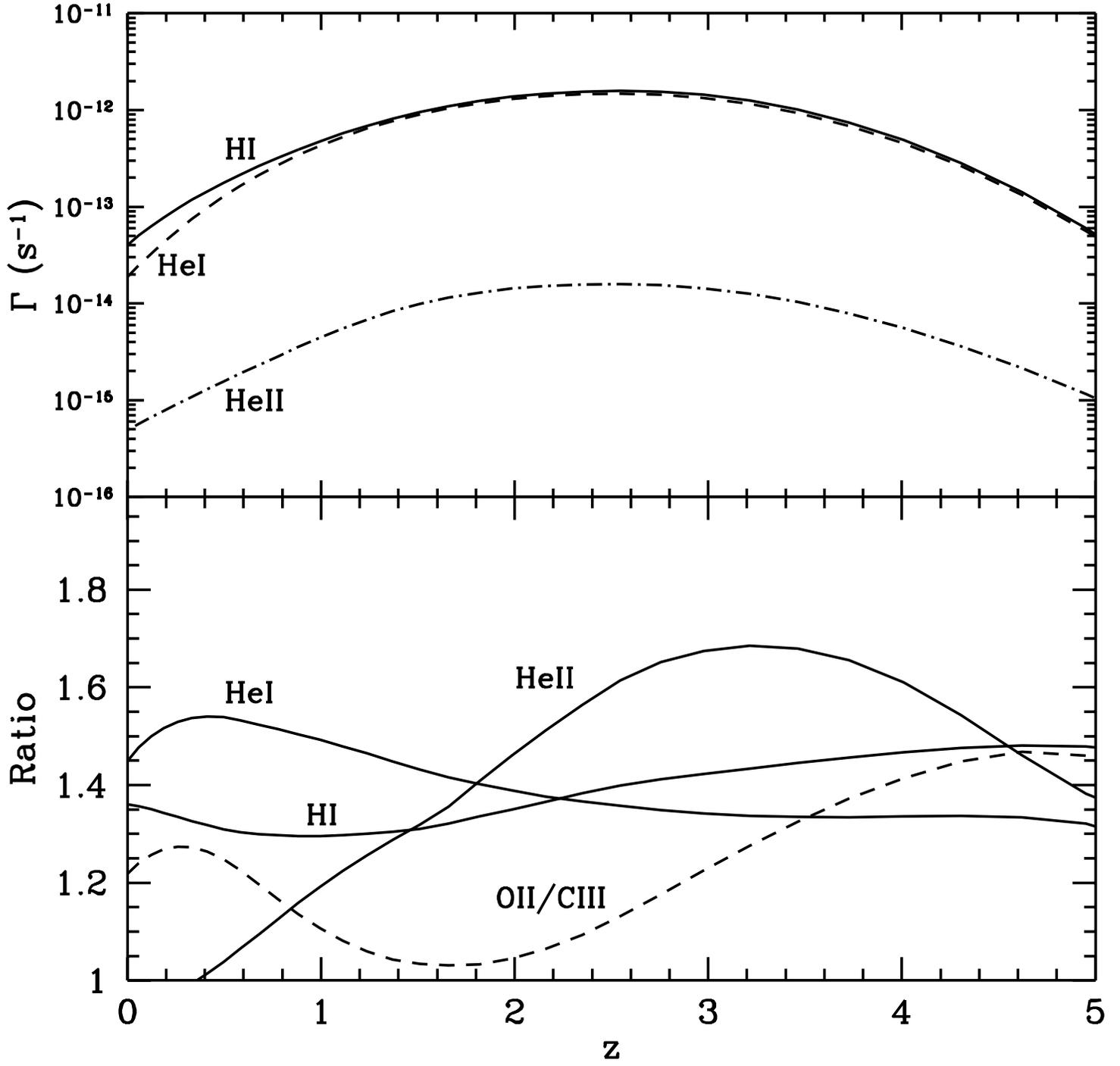

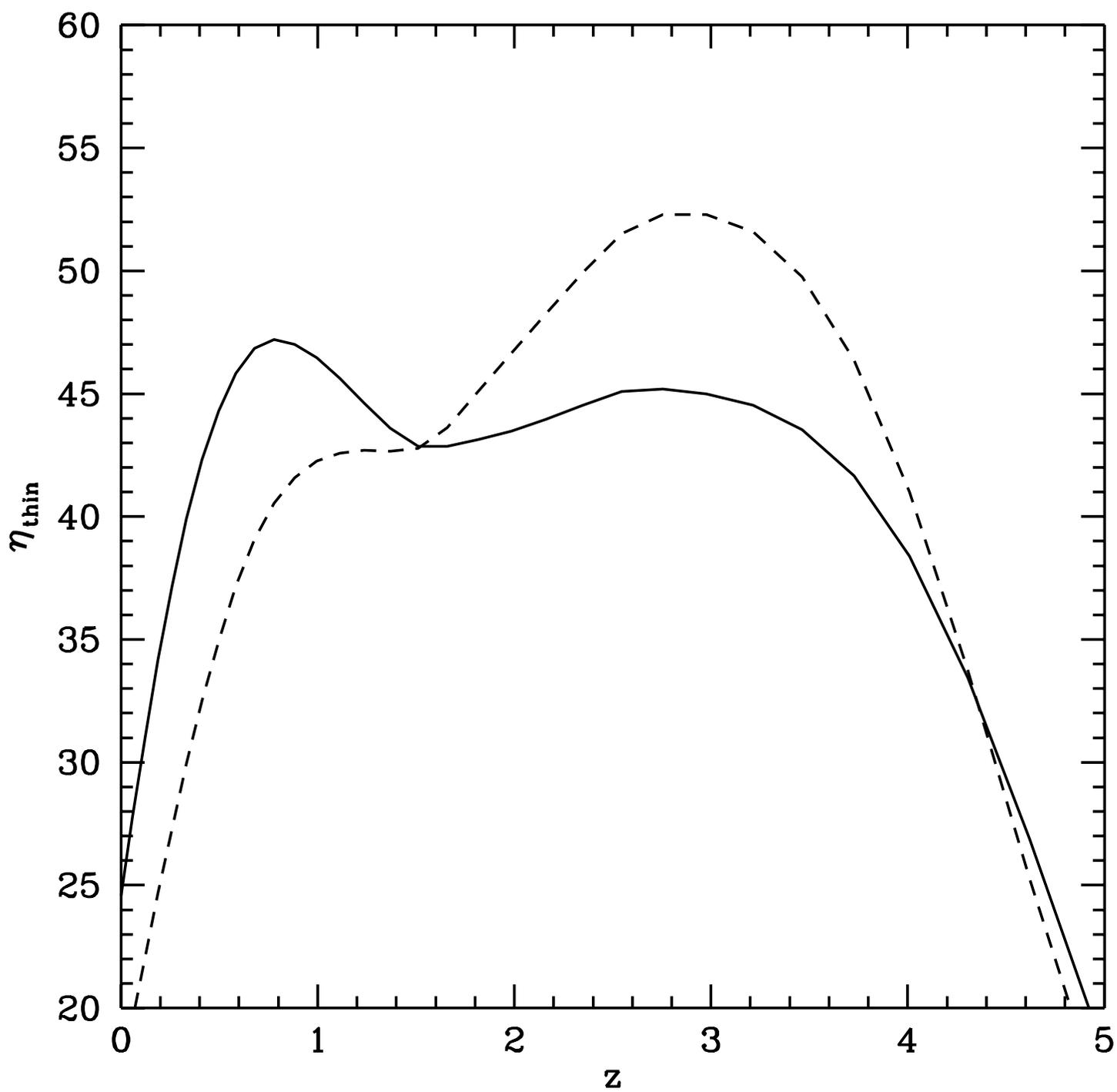

# RADIATIVE TRANSFER IN A CLUMPY UNIVERSE:
## II. THE ULTRAVIOLET EXTRAGALACTIC BACKGROUND

*Francesco Haardt*[1,2,3] *and Piero Madau*[1]

[1] Space Telescope Science Institute, 3700 San Martin Drive, Baltimore MD 21218, USA

[2] Department of Astronomy & Astrophysics, Institute of Theoretical Physics,
University of Göteborg & Chalmers University of Technology, 412 96 Göteborg, Sweden

[3] ISAS/SISSA, via Beirut 2–4, 34014 Trieste, Italy

## ABSTRACT

The integrated ultraviolet flux arising from QSOs and/or hot, massive stars in metal-producing young galaxies is likely responsible for maintaining the intergalactic diffuse gas and the Ly$\alpha$ forest clouds in a highly ionized state. The spectrum and intensity of such UV background have generally been obtained by modeling the reprocessing due to intervening material as a pure photoelectric absorption process. However, owing to the emission from radiative recombinations within the absorbing clouds, a photoionized clumpy medium could contribute substantially to the metagalactic flux. In other words, QSO absorption-line systems are sources not just sinks of ionizing photons.

We present a detailed calculation of the propagation of AGN-like ionizing radiation through the intergalactic space. We model the ionization state of absorbing clouds, and show that the universe will be more opaque above 4 Ryd than previously estimated. Singly ionized helium in Ly$\alpha$ forest clouds and Lyman-limit systems is found to be very efficient in reprocessing soft X-ray, helium-ionizing photons into ultraviolet, hydrogen-ionizing ones. We demonstrate that a significant fraction of the absorbed primary photons (emitted, e.g., by quasar sources) will be reradiated by the photoionized gas through Ly$\alpha$ line emission, two-photon continuum, and recombination continuum radiation.

In the light of new data and recent studies, we also reassess the contribution of the QSOs observed in optical surveys to the UV extragalactic background, and find that the stochastic reprocessing of quasar Lyman continuum radiation by hydrogen and helium along the line of sight will significantly affect the amplitude, spectral shape, and fluctuations properties of the metagalactic flux. In a scenario in which QSOs are the primary source of ionizing photons in the universe, the integrated H I Ly$\alpha$ emission at $z=0$ from photoionized Ly$\alpha$ clouds and Lyman-limit systems is found to be at a level of less than 5% of current observational limits on the far-UV extragalactic radiation flux. We show that $J_{912}$ increases from $\approx 10^{-23}\,{\rm ergs\,cm^{-2}\,s^{-1}\,Hz^{-1}\,sr^{-1}}$ at the present epoch to $\approx 5\times 10^{-22}\,{\rm ergs\,cm^{-2}\,s^{-1}\,Hz^{-1}\,sr^{-1}}$ at $z=2.5$. The attenuated direct quasar emission plus recombination radiation from intergalactic gas appears to provide enough hydrogen-ionizing photons to satisfy the proximity effect at large redshift. The He II /H I ratio in the diffuse intergalactic medium and the Ly$\alpha$ clouds increases from $\approx 25$ at $z=0$ to $\approx 45$ at $z=2.5$, to decrease again below 30 for $z\gtrsim 4.5$. The spectrum of the ionizing background at high redshift is shown to have a hump at energies below 40.8 eV due to redshift-smeared He II Ly$\alpha$ line and two-photon continuum emission. We propose that observations of low-ionization species such as O II in metal-line absorption systems may be able to test the presence of such a prominent feature in the UV background spectrum. We also note that, if the metagalactic flux is dominated by QSOs, the suggested steep decline of their ionizing emissivity beyond $z \sim 4$ should produce an increase in the observed rate of incidence of Ly$\alpha$ forest clouds at these redshifts relative to an extrapolation from the intermediate-$z$ regime, as observed by Williger et al. (1994).

*Subject headings*: cosmology: observations – diffuse radiation – intergalactic medium – quasars: absorption lines

# 1. INTRODUCTION

The existence of a uniformly distributed intergalactic medium (IGM), which contains the bulk of the baryonic matter of the universe in most models of structure formation, is predicted as a product of primordial nucleosynthesis. The hydrogen component of this IGM must have been highly ionized by $z \sim 5$, as the flux decrement observed on the blue side of the Ly$\alpha$ emission line of neutral hydrogen in the spectra of high redshift QSOs appears entirely consistent with the blanketing by discrete absorption lines along the line of sight (Steidel & Sargent 1987; Giallongo et al. 1994). It is widely believed (but see, e.g., Sciama 1993) that the integrated ultraviolet flux arising from QSOs and/or hot, massive stars in metal-producing young galaxies is responsible for maintaining the intergalactic diffuse gas and the Ly$\alpha$ forest clouds in a highly ionized state. Such UV background may also be responsible for the ionization of the metal-rich QSO absorption systems (Steidel 1990; Vogel & Reimers 1993) and of the hydrogen clouds located in the galactic halo (Fransson & Chevalier 1985; Ferrara & Field 1994), and for producing the sharp edges of H I disks observed in nearby spiral galaxies (Bochkarev & Sunyaev 1977; Maloney 1993; Corbelli & Salpeter 1993; Dove & Shull 1994). Photoionization by UV radiation may also inhibit the collapse of small-mass galaxies (Dekel & Rees 1987; Efstathiou 1992).

The *Hubble Space Telescope* Faint Object Camera (FOC) detection of redshifted He II 304Å absorption in the spectrum of the $z = 3.29$ quasar Q0302-003 (Jakobsen et al. 1994) has recently provided new constraints on the thermal and ionization state of the IGM, and on the spectrum of the UV metagalactic flux at early epochs. The high ratio of He II to H I which is necessary to account for the observed absorption trough requires a soft radiation field at 4 Ryd compared to 1 Ryd (Madau & Meiksin 1994; Giroux, Fardal, & Shull 1995). Although the mean spectrum of the individual sources which dominate the metagalactic flux might not be very steep, the cosmological "filtering" through material along the line of sight is known to significantly soften the background intensity. Absorption by intervening matter actually makes the universe at high redshift optically thick to Lyman continuum (LyC) photons: enough neutral hydrogen and singly ionized helium is contained in the intergalactic clumps of highly ionized gas which form the Ly$\alpha$ forest, and in the metal-line absorption systems associated with the halo regions of bright galaxies, to significantly attenuate the LyC flux from all but the nearby sources (Bechtold et al. 1987; Miralda-Escudé & Ostriker 1990; Møller & Jakobsen 1990; Madau 1991, 1992; Meiksin & Madau 1993; Zuo & Phinney 1993; Madau 1995).

This is the second paper in a series aimed at a detailed numerical study of the absorption and reprocessing of UV photons in a clumpy, photoionized universe. In Paper I (Madau 1995), we have assessed the effects of the stochastic attenuation produced by QSO absorption systems along the line of sight on the broadband colors of galaxies at cosmological distances. It is the purpose of this paper to focus on the effects of intervening discrete absorbers on the propagation of ionizing diffuse radiation. The primary motivation of this study is the realization that the integrated UV background will actually be the sum of the direct radiation from discrete sources and the emission from radiative recombinations within the absorbing clouds and the intercloud, diffuse IGM. Since the intercloud medium is optically thin to LyC photons, its contribution to the photoionizing radiation field will be negligible. We show in this paper that this will not be true, however, for the cloudy component, as a significant fraction of the absorbed primary photons will be reradiated by the photoionized gas through Ly$\alpha$ line emission, two-photon continuum, and recombination continuum radiation. The plan is as follows. In § 2 we review the basic theory of cosmological radiative transfer in a clumpy universe. In § 3 we model the physical state of absorbing clouds, as the knowledge of their ionization structure is critical to understanding the propagation of ionizing photons through intergalactic space. In § 4 we determine the contribution of recombination radiation from Ly$\alpha$ forest clouds and Lyman-limit systems to the metagalactic flux. In § 5 we numerically compute the spectrum of the background radiation field as a function of redshift, assuming that the quasars detected in optical surveys are the primary source of LyC photons in the universe. We derive expressions for the hydrogen and helium ionization rates, and discuss the implications of our results for the Gunn-Peterson test, the proximity effect, the He II Ly$\alpha$ cosmic opacity, the fluctuations of the metagalactic ionizing flux at high redshift, and the properties of the Ly$\alpha$ forest clouds and metal-line absorption systems. We briefly summarize our conclusions in § 6. In Appendix A we provide some fitting formulae to the recombination rate coefficients, while in Appendix B we present useful analytical approximations to the effective cosmic opacity and the recombination emissivity. Finally, in Appendix C we discuss a numerical algorithm for a fast iterative solution to the equation of radiative transfer. Readers who are interested in the main results and not in the technical details of the model may wish to proceed directly to § 5 at this point.

## 2. BASIC THEORY

### 2.1. Cosmological Radiative Transfer

The radiative transfer equation in cosmology describes the evolution in time of the specific intensity $J$ (in units of ergs cm$^{-2}$ s$^{-1}$ Hz$^{-1}$ sr$^{-1}$) of a diffuse radiation field (e.g., Peebles 1993):

$$\left(\frac{\partial}{\partial t} - \nu \frac{\dot{a}}{a} \frac{\partial}{\partial \nu}\right) J = -3\frac{\dot{a}}{a} J - c\kappa J + \frac{c}{4\pi}\epsilon, \qquad (1)$$

where $a$ is the cosmological scale parameter, $c$ the speed of the light, $\kappa$ the continuum absorption coefficient per unit length along the line of sight, and $\epsilon$ is the proper space-averaged volume emissivity. Integrating equation (1) and averaging over all lines of sight yields the mean specific intensity of the radiation background at the observed frequency $\nu_o$, as seen by an observer at redshift $z_o$:

$$J(\nu_o, z_o) = \frac{1}{4\pi} \int_{z_o}^{\infty} dz \, \frac{d\ell}{dz} \frac{(1+z_o)^3}{(1+z)^3} \epsilon(\nu, z) e^{-\tau_{eff}(\nu_o, z_o, z)}, \qquad (2)$$

where $\nu = \nu_o(1+z)/(1+z_o)$,

$$\frac{d\ell}{dz} = \frac{c}{H_0}(1+z)^{-2}(1+2q_0 z)^{-1/2} \qquad (3)$$

is the line element in a Friedmann cosmology, $H_0 = 50 h_{50}$ km s$^{-1}$ Mpc$^{-1}$ the Hubble constant, and $q_0$ the deceleration parameter.

### 2.2. Intergalactic Absorption

The *effective optical depth* $\tau_{eff}$ due to discrete absorption systems is defined as

$$\tau_{eff} = -\ln(\langle e^{-\tau} \rangle), \qquad (4)$$

where $\langle e^{-\tau} \rangle$ is the average transmission over all lines of sight. For Poisson-distributed clouds we have (Paresce, McKee, & Bowyer 1980)

$$\tau_{eff}(\nu_o, z_o, z) = \int_{z_o}^{z} dz' \int_{0}^{\infty} dN_{\rm HI} \frac{\partial^2 N}{\partial N_{\rm HI} \partial z'}(1 - e^{-\tau}), \qquad (5)$$

where $\partial^2 N/\partial N_{\rm HI} \partial z'$ is the redshift and column density distribution of absorbers along the line of sight. The continuum optical depth through an individual cloud is (assuming a pure hydrogen-helium gas)

$$\tau(\nu') = N_{\rm HI}\sigma_{\rm HI}(\nu') + N_{\rm HeI}\sigma_{\rm HeI}(\nu') + N_{\rm HeII}\sigma_{\rm HeII}(\nu'), \qquad (6)$$

where $\nu' = \nu_o(1+z')/(1+z_o)$, and $\sigma_{\rm HI}, \sigma_{\rm HeI}$, and $\sigma_{\rm HeII}$ are the hydrogen and helium photoionization cross-sections. When $\tau \ll 1$, $\tau_{eff}$ becomes equal to the mean optical depth. In the opposite limit, the obscuration is picket fence-type, and the effective optical depth becomes equal to the mean number of optically thick systems along the line of sight.

Two classes of absorbers are known to dominate the cosmic LyC opacity, with the Ly$\alpha$ forest clouds being the most numerous. These have neutral column densities ranging from $\sim 10^{13}$ to maybe $10^{17}$ cm$^{-2}$, close to primordial metal abundances, and evolve rapidly between $1.7 < z < 4$ (Murdoch et al. 1986). The distribution of the number of clouds as a function of H I column roughly follows a single power-law $dN/dN_{\rm HI} \propto N_{\rm HI}^{-\beta}$, with $\beta \sim 1.5$ (Tytler et al. 1995a; Tytler 1987). The redshift evolution of the rarer Lyman-limit systems (log $N_{\rm HI} > 17.2$), which are optically thick to radiation with energy greater than 1 Ryd, is much weaker over the range $0.7 < z < 3.6$ (Storrie-Lombardi et al. 1994; Sargent, Steidel, & Boksenberg 1989). About 2/3 of the Lyman-limit absorbers are associated with heavy-element lines.

To facilitate a comparison with previous estimates of the UV background (Miralda-Escudé & Ostriker 1990; Madau 1991, 1992; Miralda-Escudé & Ostriker 1992; Meiksin & Madau 1993; Giroux & Shapiro 1995), we assume in the following a constant value of $\beta = 1.5$ over 7 decades in $N_{\rm HI}$. For the redshift and column density distribution of intervening absorbers we take

$$\frac{\partial^2 N}{\partial N_{\rm HI} \partial z} = \begin{cases} N_0 N_{\rm HI}^{-1.5}(1+z)^{\gamma} & (10^{13} < N_{\rm HI} < 1.59 \times 10^{17} \text{ cm}^{-2}); \\ 5.4 \times 10^7 N_{\rm HI}^{-1.5}(1+z)^{1.55} & (1.59 \times 10^{17} < N_{\rm HI} < 10^{20} \text{ cm}^{-2}), \end{cases} \qquad (7)$$

where $N_0 = 2.9 \times 10^7$ and $\gamma = 2.46$ for $z > 1.7$. With the above normalization, there are $dN/dz = 4.3(1+z)^{2.46}$ Ly$\alpha$ forest clouds with rest equivalent width $W > 0.32$ Å (corresponding to $N_{\rm HI} > 1.8 \times 10^{14}$ cm$^{-2}$ for an average Doppler width of 28 km s$^{-1}$), as measured at high redshift by Press & Rybicki (1993). At $z \sim 0$, the number of Ly$\alpha$ forest line is $\sim 5$ times larger than expected from a simple extrapolation of the high redshift evolution (Bahcall et al. 1991; Morris et al. 1991). For $z < 1.7$, we therefore take $\gamma = 1.0$ and normalize imposing continuity at $z = 1.7$. This gives $\sim 18$ lines per unit redshift with $W > 0.32$ Å at the present epoch, consistent with the findings of Bahcall et al. (1993). At high-$z$, the model adopted will provide about the same photoelectric opacity as in model A2 of Miralda-Escudé & Ostriker (1990), and in the medium attenuation model MA of Meiksin & Madau (1993). It is fair to point out that deviation from a single power-law in column density could be significant. In particular, there is very limited information on Ly$\alpha$ forest absorbers with $10^{16} \lesssim N_{\rm HI} \lesssim 10^{17}$ cm$^{-2}$. These high column density clouds will dominate the Ly$\alpha$ forest contribution to the effective photoelectric optical depth, making its value rather uncertain. Meiksin & Madau (1993) have suggested that the published data may actually indicate an underdensity of Ly$\alpha$ with columns $N_{\rm HI} > 10^{15}$ cm$^{-2}$. Giallongo et al. (1993), Petitjean et al. (1993), and Kulkarni et al. (1995) have also reported evidence for a break in the $N_{\rm HI}$ power-law at $\log N_{\rm HI} \gtrsim 15$. This deficit could result in a significant reduction of the forest continuum opacity, and therefore in an increased background ionizing flux. We will discuss this possibility in more details in § 5.

In the following, we will also neglect the photoelectric opacity associated with a uniformly distributed IGM. This is consistent with the lack of a H I Gunn-Peterson absorption trough in the spectra of QSOs at large redshift (e.g., Giallongo et al. 1994). Note, however, that if the recent *Hubble Space Telescope* observation of a He II Ly$\alpha$ trough in the spectrum of quasar Q0302-03 (Jakobsen et al. 1994) were to require a Gunn-Peterson optical depth $\gtrsim 10$, the He II LyC opacity of a diffuse IGM component would be significant (Madau & Meiksin 1994).

## 3. THE IONIZATION STATE OF QUASAR ABSORPTION SYSTEMS

A self-consistent calculation of the effective optical depth in equation (5) requires a detailed understanding of the ionization state of intervening material. In particular, the amount of He I and He II which is present along the line of sight must be inferred by modeling the physical conditions in the absorbing clouds themselves. This will be done in this section. We will assume the gas to be in ionization equilibrium with the background UV radiation. Deviation from equilibrium will be small, of order the ratio of the ionization timescale to the Hubble time, $t_{ion}/t_{\rm H} \sim 6 \times 10^{-6} J_{912,-21}^{-1}[(1+z)/4]^{3/2} h_{50}$, where the background intensity is measured at the hydrogen Lyman edge in units of $10^{-21}$ ergs cm$^{-2}$ s$^{-1}$ Hz$^{-1}$ sr$^{-1}$.

In principle, the temperature $T$ of the absorbing gas should also be computed self-consistently, by balancing photoionization heating with free-bound and free-free radiative losses, collisional excitation losses, and Compton cooling losses from scattering off the microwave background. The temperature fully ionized gas may reach depends, however, on the poorly known cloud density. For $n_{\rm H} \gtrsim 4 \times 10^{-4}$ cm$^{-3}$, collisional excitation of neutral hydrogen is the dominant cooling mechanism, and the cloud temperature increases with decreasing density, to drop again at lower densities when Compton cooling starts to dominate (Meiksin 1994). Moreover, when $n_{\rm H} \lesssim 10^{-4}$ cm$^{-3}$ (the typical density of a spherical Ly$\alpha$ cloud with neutral hydrogen column $10^{14}$ cm$^{-2}$ and diameter 100 kpc in equilibrium with a radiation field of intensity $J_{912,-21} \sim 1$), the cooling timescale is longer than the age of the universe at the redshift of interest, and the gas will retain a memory of the initial conditions. The contraction or expansion of the gas toward hydrostatic equilibrium will further affect its thermal properties. On the other hand, we can estimate the temperature of the absorbing gas from the measurements of Ly$\alpha$ line profiles in high resolution, high signal-to-noise spectra. For simplicity, we shall assume here a fixed temperature of $T = 2.5 \times 10^4$ K. This is the value derived from modeling the broader Ly$\alpha$ absorption lines as blends of $\approx 3$ clouds each with thermal width $(2k_{\rm B}T/m_{\rm H})^{1/2} \simeq 20$ km s$^{-1}$, and a velocity dispersion of $\approx 11$ km s$^{-1}$ (Hu et al. 1995). We will also assume that intergalactic material acts only as a "filter" of external background photons emitted from quasars and/or star-forming galaxies, i.e., there are no "local" discrete sources of UV radiation within individual clouds. Although the photoionization equilibrium of a H and He gas is standard textbook topic, we are applying it in such an unusually wide range of parameters (about 7 decades in $N_{\rm HI}$ and nearly 2 decades in $J_{912}$) that it is worth stating the basic equations and approximations used.

### 3.1. Clouds Optically Thin To LyC Radiation

If we define $Y_{\rm HI} \equiv N_{\rm HI}/N_{\rm H}$, $Y_{\rm HeI} \equiv N_{\rm HeI}/N_{\rm He}$ and $Y_{\rm HeII} \equiv N_{\rm HeII}/N_{\rm He}$, the steady state ionization equilibrium equations can be written as

$$\begin{aligned} 1 - Y_{\rm HI} &= Y_{\rm HI} I_{\rm HI} \\ Y_{\rm HeII} &= Y_{\rm HeI} I_{\rm HeI} \quad , \\ 1 - Y_{\rm HeI} - Y_{\rm HeII} &= Y_{\rm HeII} I_{\rm HeII} \end{aligned} \quad (8)$$

where

$$I_{\rm HI} \equiv \frac{\Gamma_{\rm HI}}{n_e \alpha_{\rm HI}(T)}. \quad (9)$$

Here,

$$\Gamma_{\rm HI} = \int_0^\infty d\nu \, 4\pi \frac{J(\nu)}{h\nu} \sigma_{\rm HI}(\nu) \quad (10)$$

is the photoionization rate per hydrogen atom, $\alpha_{\rm HI}$ the (Case A) recombination rate coefficient to all levels of hydrogen, $n_e$ the electron number density, and $J$ the background radiation intensity. Analogous expressions are valid for $I_{\rm HeI}$ and $I_{\rm HeII}$. Collisional ionizations have been neglected: we will discuss in § 3.4 the validity of this approximation. The radiative recombination coefficients for hydrogenic ions are taken from Arnaud & Rothenflug (1985), while those for neutral helium are taken from Black (1981). The H I and He II photoionization cross-sections are adopted from Osterbrock (1989). For the photoionization cross-section of neutral helium we used the following fit (to within a few percent up to 1 keV) of the results of Reilman & Manson (1979):

$$\sigma_{\rm HeI} \simeq \frac{0.694 \times 10^{-18}}{E^{1.82} + E^{3.23}} \; {\rm cm}^2, \quad (11)$$

for $E > 0.246$, where $E$ is the photon energy in units of 100 eV.

Notice that, while He III recombines for $J_{228,-24} \lesssim n_{\rm H,-4}$, hydrogen is mostly ionized for $J_{912,-24} \gtrsim 0.1 \, n_{\rm H,-4}$. Thus, in the case of a QSO-dominated background, the typical conditions of the Ly$\alpha$ clouds assure that most of the gas is completely ionized. In this case we can write

$$\tau \approx N_{\rm HI}(\sigma_{\rm HI} + \eta_{thin} \sigma_{\rm HeII}), \quad (12)$$

where

$$\eta_{thin} \equiv \left.\frac{N_{\rm HeII}}{N_{\rm HI}}\right|_{thin} = \frac{n_{\rm He}}{n_{\rm H}} \frac{(1 + I_{\rm HI})I_{\rm HeI}}{1 + I_{\rm HeI}(1 + I_{\rm HeII})} \approx \frac{1}{12}\left(\frac{I_{\rm HI}}{I_{\rm HeII}}\right), \quad (13)$$

assuming a helium to hydrogen cosmic abundance ratio equal to 1/12.

The solution of the ionization equilibrium given above neglects any kind of radiative transfer within the individual clouds, and has been extensively used in previous investigations (Miralda-Escudé & Ostriker 1990, 1992; Madau 1991, 1992; Giroux & Shapiro 1995). These authors have argued that equations (12) and (13) might describe reasonably well the overall transmission properties of the universe even when optically thick clouds are present. The reason is that, for AGN-like or softer spectra, the ratio $\eta_{thin} \sim 1.8 J_{912}/J_{228}$ is much greater than unity: this implies that clouds which are about to become optically thick in the He II LyC are still optically thin in the H I LyC. To be more specific, let us denote with $N_{\rm HI}(4\,{\rm Ryd})$ the H I column at which He II self-shielding sets in. Clouds with $N_{\rm HI} = N_{\rm HI}(4\,{\rm Ryd})$ have optical depth of unity at 4 Ryd, or, equivalently, an He II column $N_{\rm HeII}(4\,{\rm Ryd}) \equiv \sigma_{\rm HeII}^{-1}(4\,{\rm Ryd}) = 6.35 \times 10^{17}\,{\rm cm}^{-2}$. (Similarly, self-shielding at 1 Ryd sets in above $N_{\rm HI}(1\,{\rm Ryd}) \equiv \sigma_{\rm HI}^{-1}(1\,{\rm Ryd}) = 1.59 \times 10^{17}\,{\rm cm}^{-2}$.) While our estimate of the singly ionized helium to neutral hydrogen ratio from equation (13) is then quite accurate for all values of $N_{\rm HI}$ less than $N_{\rm HI}(4\,{\rm Ryd})$, it is clearly inadequate at larger H I columns. In this regime, the absorption of He II LyC photons within the clouds must be taken into account in calculating the expected ionization structure. It turns out, however, that the contribution of clouds with $N_{\rm HI} > N_{\rm HI}(4\,{\rm Ryd})$ to the effective optical depth of the universe at 4 Ryd is more or less independent of their actual He II column. This is because the exact value of $\tau$ for $\tau$ greater than a few is irrelevant in equation (5), as most of the photons are removed anyway. For a column density distribution $dN/dN_{\rm HI} \propto N_{\rm HI}^{-1.5}$ like the one given in equation (7), most of the effective opacity at frequency $\nu$ actually comes from absorbers with $\tau(\nu) \sim 1$.

We have found that, although the optically thin formula correctly estimates the cosmic transmission at (or just above) the He II Lyman edge, it breaks down when $\lambda \ll 228\,{\rm \AA}$. The problem is that, due to the wavelength dependence of LyC absorption, absorbers which are just optically thick for these penetrating photons [$\tau(\lambda \ll 228\,{\rm \AA}) \sim 1$] are very optically thick at 4 Ryd. In these clouds, the ratio $N_{\rm HeII}/N_{\rm HI}$ is generally much higher than the value estimated from equation (13), since He III almost completely recombines (see below). The trasmittance of photons above 4 Ryd actually drops to zero for clouds with $N_{\rm HI} > N_{\rm HI}(4\,{\rm Ryd})$. *The net effect is that the Universe will be more opaque above 4 Ryd than previously estimated.* In the next section we will provide a more accurate, albeit still approximate, radiative transfer solution for the actual ionization state of individual optically thick absorbers.

### 3.2. Clouds Optically Thick to LyC Radiation

We shall consider a pure H/He semi-infinite slab. The ionization equilibrium equation for the neutral hydrogen density at depth $R$ within the slab is

$$\alpha_{\rm HI}(T) n_e n_{\rm HII} = n_{\rm HI} \int_{\nu_L}^{\infty} d\nu\, \sigma_{\rm HI}(\nu) \frac{4\pi J(\nu)}{h\nu} e^{-\tau(\nu)} + \int_{\nu_L}^{\infty} d\nu\, p_{abs}(\nu) \frac{\epsilon_r(\nu)}{h\nu}, \qquad (14)$$

where $h\nu_L = 1\,{\rm Ryd}$,

$$\tau(\nu) = \int_0^R dr\, [n_{\rm HI}\sigma_{\rm HI}(\nu) + n_{\rm HeI}\sigma_{\rm HeI}(\nu) + n_{\rm HeII}\sigma_{\rm HeII}(\nu)], \qquad (15)$$

and the term $p_{abs}$ is the fraction of recombination continuum radiation with emissivity $\epsilon_r$ which is absorbed within the slab. Analogous expressions can be written for the helium ions. Since the recombination emissivity is strongly peaked at threshold, we can approximate the last term in equation (14) as

$$\int_{\nu_L}^{\infty} d\nu\, p_{abs}(\nu) \frac{\epsilon_r(\nu)}{h\nu} \approx p_{abs}(1\,{\rm Ryd}) \alpha_{1^2S} n_e n_{\rm HII}, \qquad (16)$$

where $\alpha_{1^2S}$ is the recombination coefficient to the ground level of hydrogen (see Appendix A). For simplicity, and because of the steep dependence of the photoionization cross-section with energy, we have neglected local helium recombination radiation as a source of hydrogen-ionizing photons.

Note, however, that helium recombination photons will be more effective in ionizing hydrogen as they escape into the intergalactic space, and are redshifted by the cosmological expansion (cf § 4). We shall take

$$p_{abs}(1\,\text{Ryd}) = 1 - \frac{1 - e^{-\tau(1\,\text{Ryd})}}{\tau(1\,\text{Ryd})}. \tag{17}$$

Equation (17) is strictly valid only in regions of constant ion and electron density. [1] It is possible to show that, for the conditions typically encountered, the product $n_e n_{\text{HII}}$ is indeed fairly constant even throughout absorbers which are very optically thick (log $N_{\text{HI}} \approx 19$) to hydrogen LyC radiation. This is not true for helium, however, as He III recombines faster than hydrogen, and the radiation intensity at the He II edge is in general smaller than the intensity at the H I ionization edge. The He III number density in clouds which are optically thick at 4 Ryd is constant up to the He II photosphere, where $n_{\text{HeII}} \sigma_{\text{HeII}}(4\,\text{Ryd}) R \approx 1$, then drops sharply by more than an order of magnitude as He III starts to recombine. Since most of the reemission and absorption of photons above 4 Ryd occurs at the He II photosphere, and $\tau(h\nu \geq 4\,\text{Ryd}) \approx \tau_{\text{HeII}}(\nu)$, the absorbed fraction of He II recombination radiation can be written as

$$p_{abs}(4\,\text{Ryd}) \approx 1 - \frac{1 - e^{-\tau^*}}{\tau^*}, \tag{18}$$

where

$$\tau^* \equiv \min[\int_0^R dr\, n_{\text{HeII}}(r)\sigma_{\text{HeII}}(4\,\text{Ryd}),\, 1]. \tag{19}$$

Therefore, while only a fraction $\tau^{-1}(1\,\text{Ryd})$ of H I LyC photons will escape from clouds which are very optically thick at 1 Ryd, at least 60% of the He II LyC photons produced in individual systems will always be able to leak out into the IGM.

We can now solve for the $n_{\text{HI}}$, $n_{\text{HeI}}$, and $n_{\text{HeII}}$ number densities as a function of $R$. We obtain

$$\frac{n_{\text{HI}}}{n_{\text{H}}} = \frac{n_e \tilde{\alpha}_{\text{HI}}}{\int_{\nu_L}^{\infty} d\nu\, \frac{4\pi J(\nu)}{h\nu}\sigma_{\text{HI}}(\nu)e^{-\tau(\nu)} + n_e \tilde{\alpha}_{\text{HI}}}, \tag{20}$$

where $\tilde{\alpha}_{\text{HI}}(\tau) \equiv \alpha_{\text{HI}} - p_{abs}(1\,\text{Ryd})\alpha_{1^2S}$. Again, similar expressions can be written for the $n_{\text{HeI}}/n_{\text{He}}$ and $n_{\text{HeII}}/n_{\text{He}}$ number ratios. Finally, the H I, He I, and He II column densities can be derived by integrating through the slab.

### 3.3. Simple Approximations

An iterative solution to the radiative transfer equation (2) which included a detailed computation of the ionization structure of individual absorbers would be very time consuming. In this section we develop a simple approximation to the exact form of the function $\eta \equiv$ He II /H I which is capable of reproducing the full absorption properties of intergalactic material.

---

[1] Assuming a spherical geometry for the clouds would change eqs. (16) and (17), allowing slightly more photons to escape for a given optical depth. For $\tau \gg 1$, $p_{abs} \approx 1$ independent of geometry. When $\tau \ll 1$, the difference between a slab and a sphere can be significant, but in this case $p_{abs} \ll 1$ anyway, and its exact value is irrelevant in the definition of $\tilde{\alpha}$.

The asimptotic limit for low $N_{\rm HI}$ is readily obtained in the optically thin regime (see eq. [13]). The self-shielding of He II LyC radiation becomes important as $N_{\rm HI}$ approaches the value

$$N_{\rm HI}(4\,{\rm Ryd}) = 4\,N_{\rm HI}(1\,{\rm Ryd})\eta_{thin}^{-1}. \tag{21}$$

At about twice this value He III recombines and the He II column rises sharply, while hydrogen is still mostly ionized. The He II to H I column density ratio quickly reaches its maximum value,

$$\eta_{max} \simeq \frac{n_{\rm He}}{n_{\rm H}} \frac{0.37\,\Gamma_{\rm HI} + n_e \tilde{\alpha}_{\rm HI}(\tau=1)}{n_e \tilde{\alpha}_{\rm HI}(\tau=1)}, \tag{22}$$

where the factor 0.37 ($\simeq 1/e$) takes into account the partial self-shielding of neutral hydrogen, $\Gamma_{\rm HI}$ is the photoionization rate in the limit of an optically thin cloud, $\tilde{\alpha}_{\rm HI}(\tau=1) = \alpha_{\rm HI} - 0.37\alpha_{1^2S}n_e$, and we have assumed a negligible fraction of neutral helium. It is straightforward to write down a more general expression including the He I component. The contribution of He I to the overall opacity of the universe is fully taken into account in our numerical calculations. This is found to be negligible except at the very highest redshift, $z \gtrsim 5$, where the quasar emissivity drops and helium starts to recombine (see § 5 below). Finally, at even higher H I columns, $N_{\rm HI} >$ several $\times 10^{17}\,{\rm cm}^{-2}$, both H II and He II recombine. The He II to H I ratio becomes a decreasing function of H I and finally reaches the asymptotic limit $\eta_{thick} = 0$ of a completely neutral gas.

We have found that the exact form of $\eta(N_{\rm HI})$ can be well approximated by a step-like function, as shown in Figure 1 for different values of the background intensity, an assumed temperature of $T = 2.5 \times 10^4$ K, and a hydrogen gas density of $n_{\rm H} = 2 \times 10^{-4}\,{\rm cm}^{-3}$ (a larger $n_{\rm H}$ would decrease the value of $\eta_{max}$, thereby decreasing the cosmological effective opacity above 54.4 eV). The drop at large hydrogen columns is well modeled by assuming $\eta \propto 1/N_{\rm HI}$ for $N_{\rm HI} \gtrsim 6 \times 10^{17}\,{\rm cm}^{-2}$. A similar approximation has been used for the neutral helium component. Note that, in order to take into account the fact that both faces of the slab are actually illuminated by the metagalactic flux, we have associated the value of $\eta$ at a given $R$ to a cloud with column density equal to $2N_{\rm HI}(R)$. Figure 2 shows, again as a function of H I column, the absorption probability through an individual cloud, $1 - \exp(-\tau)$, for various ionizing photon energies. We compare the exact value of this probability function with those obtained from the "step-like approximation" and the optically thin, $\eta = {\rm const} = \eta_{thin}$, limit used in previous investigations. When integrated over the actual column density distribution, the step-like approximation gives typical errors of a few percent, with a maximum value $\sim 20\%$ for photon energies $\gtrsim 200$ eV. However, at such short wavelengths, the effective optical depth of the universe is $\lesssim 1$, and the percental error in computing the specific background intensity $J$ ($\propto e^{-\tau_{eff}}$) is a factor $\sim \tau_{eff}$ smaller.

Since the integration over $N_{\rm HI}$ in equation (5) can be done analytically over ranges of constant $\eta$ and for $\eta \propto 1/N_{\rm HI}$, the step-like approximation for the ionization structure of individual absorbers along the line of sight reduces significantly the computational time. This will allow a fast computation of the contribution of recombination radiation to the specific intensity (see Appendix B). Different possibilities for the nature of the ionizing sources and uncertainties in the column and redshift distribution of intervening absorbers can then be quickly explored.

### 3.4. Importance of Collisional Effects

In the relevant range of densities for the Ly$\alpha$ clouds and the Lyman-limit systems, $n_{\rm H} \approx 10^{-4}$–$10^{-2}\,{\rm cm}^{-3}$, collisional ionizations are always negligible for the helium component because of the large ionization potentials. For hydrogen, collisional ionization can be important when the photoionization rate is low, i.e. for low background intensities or in the core of optically thick clouds. A related issue concerns the importance of collisionally excited line emission, notably H I Ly$\alpha$, relative to recombination radiation. Since collisional processes are strong functions of the gas temperature, it is important to check that they are indeed negligible in the relevant regime. We have therefore solved the equation of thermal equilibrium for a gas slab immersed in a background radiation field. In Figure 3 (upper panel) we show, for three different background intensities, the average temperature of the cloud as a function of H I column density. For illustrative purposes, and to show the dependence on the assumed gas density, the total hydrogen density is taken to be $2 \times 10^{-4}$ cm$^{-3}$ for the Ly$\alpha$ forest clouds ($N_{\rm HI} < 1.59 \times 10^{17}$ cm$^{-2}$), and $4 \times 10^{-3}$ cm$^{-3}$ for the thicker Lyman-limit systems. Heating is provided by photoionization of hydrogen and helium, while recombinations and collisional excitation of neutral hydrogen dominate the cooling term.

Figure 3 (lower panel) also shows the fractional population of the $n = 2$ level of atomic hydrogen due to recombinations of photoionized H II ions, collisional excitations from the ground state, and recombinations of collisionally ionized H II ions, for clouds of different H I columns. The last contribution is never important, assuring that collisional ionizations can be neglected in the computation of the ionization structure of absorbing clouds. This is also seen in Figure 1, where the dotted lines depict the derived He II to H I ratio including collisional effects. Deviations from the values computed in § 3.3 by assuming a purely photoionized isothermal slab are indeed small. It is fair to note, however, that a fraction $\sim 20 - 30\%$ of H I Ly$\alpha$ and two photon decay emissions could still be due to radiative de-excitations of collisionally excited atoms. This contribution has not been included in our numerical calculations of the metagalactic flux.

## 4. RECOMBINATION RADIATION FROM QUASAR ABSORPTION SYSTEMS

The attenuation due to the accumulated H I and He II LyC absorption by Ly$\alpha$ forest clouds and Lyman-limit systems causes a large reduction of the background ionizing flux. However, a significant fraction of the absorbed photons is reradiated by the photoionized gas in the ultraviolet band. In this section we will show that diffuse radiation from intergalactic material will in fact dominate the cosmic ionizing emissivity at characteristic frequencies. Candidates for the source of the metagalactic flux includes AGNs (see § 5) and star-forming galaxies, but the precise nature of the background is not important in this context.

The dominant emission from a highly photoionized, dust-free gas which contains, by number, 92% of hydrogen, 8% of helium, and traces of metals, at a temperature of 2–4$\times 10^4$ K, is due to radiative recombinations and collisional excitations of the H I and He II Ly$\alpha$ lines at 1216 and 304 Å, respectively. As only Lyman series lines of He II ions can significantly affect the ionization state of the universe, and since in the considered range of densities collisionally excited He II line emission is always negligible, we have taken into account only radiation arising from radiative recombination processes. Under the typical physical conditions encountered, the hydrogen and helium ions in the absorbing medium are in their ground state as recombinations to excited levels are followed by radiative decays to the ground level. Among direct recombinations and radiative decays, only few processes are important as sources of far-ultraviolet photons. In the following we analyze them in turn.

*a) He II processes*    Direct captures of free electrons to the $n^2L$ levels of He II are an ob-

vious source of ionizing photons. As the resulting continuum radiation is strongly peaked at frequency $\nu_{th} = Z^2/n^2 \nu_L$, direct captures to the ground level will produce a significant amount of He II ionizing photons, while captures to the $2^2L$ level will radiate a Balmer continuum just above $\nu_L$. Within clouds which are optically thick to resonant scattering in the lines of the Lyman series, all helium recombinations eventually populate the $2^2S$ and $2^2P$ levels and are converted into two-photon continuum and Ly$\alpha$ 304 Å photons. In the absence of metals and dust the Ly$\alpha$ photons will diffuse into the wings and finally escape. Most He II recombination radiation will be emitted from clouds with $N_{\rm HI} \sim N_{\rm HI}(4\,{\rm Ryd}) \sim 5 \times 10^{16}$ cm$^{-2}$, as these are the dominant absorbers of photons above 4 Ryd, and will escape into the intergalactic space without appreciable destruction by local H I absorption.

*b) H I processes*   Although only free-bound captures to the ground state of hydrogen are able to generate photons above 1 Ryd, we have also computed the contribution of H I Ly$\alpha$ and two-photon continuum to the near-ultraviolet background. As in the case of He II processes, we have taken into account only radiation arising from radiative recombinations. As discussed in § 3.4, these account for at least 70–80% of the total hydrogen Ly$\alpha$ and two-photon continuum emission.

*c) Emissivity from Radiative Recombinations*   To summarize, our calculations will include the emission from Ly$\alpha$ forest clouds and Lyman-limit systems due to H I and He II Ly$\alpha$ recombination radiation, two-photon continuum, and free-bound captures to the ground and (for He II only) to the $2^2L$ levels. We have ignored He I recombination radiation because of the smallness of the $N_{\rm HeI}/N_{\rm HeII}$ and $N_{\rm HeI}/N_{\rm HI}$ ratios encountered in typical situations. Note that this will be a poor approximation at the very highest redshift, $z \gtrsim 5$, where the abundance of neutral helium increases as the space density of observed quasars drops.

In general, the specific recombination emissivity associated with a given atomic transition can be expressed as

$$\epsilon_r(\nu, z) = h\nu\, f(\nu)\, B(z, \nu)\, \frac{dz}{d\ell}, \qquad (23)$$

where

$$B(z, \nu) = \int_0^\infty dN_{\rm HI}\, \frac{\partial^2 N}{\partial N_{\rm HI} \partial z}\, \frac{\alpha^{eff}}{\tilde\alpha}\, p_{em}(\nu) \int_{Z^2 \nu_L}^\infty d\nu'\, \frac{4\pi J(\nu', z)}{h\nu'}\, w_{abs}(\nu'). \qquad (24)$$

This formula simply states that the rate of recombination photons which escape into the intergalactic space is proportional to the number of photons absorbed per second at any given redshift by the relevant atom or ion, times the fraction of recombinations which lead to the radiative transition under consideration (the ratio $\alpha^{eff}/\tilde\alpha$, where $\alpha^{eff}$ is the effective recombination coefficient, Osterbrock 1989), times the fraction $p_{em}$ of the emitted photons which are able to escape from an individual cloud. The fraction of incident photons which are absorbed by a given ion $X$, having optical depth $\tau_X$, can be written as

$$w_{abs}(\nu') = \int d\tau_X(\nu') e^{-\tau(\nu')} \approx [1 - e^{-\tau(\nu')}]\, \frac{\tau_X(\nu')}{\tau(\nu')}. \qquad (25)$$

The last equality is correct as long as the He II to H I and He I to H I ratios are constant through the slab, and it is easily generalized if these ratios vary in a step-like manner instead. In analogy with equation (17) and (18), the fraction of He II LyC photons which can escape from a cloud is

$$p_{em}(\nu > 4\nu_L) \simeq \begin{cases} \frac{1-\exp[-\tau(\nu)]}{\tau(\nu)} & [N_{\rm HI} \lesssim N_{\rm HI}(4\,{\rm Ryd})]; \\ \frac{1-\exp[-\tau(\nu)/\tau(4\nu_L)]}{[\tau(\nu)/\tau(4\nu_L)]} & [N_{\rm HI} \gtrsim N_{\rm HI}(4\,{\rm Ryd})]. \end{cases} \quad (26)$$

Note that for energies $\gtrsim 54.4$ eV we have that $\tau(\nu) \simeq \tau_{\rm HeII}(\nu)$. Only hydrogen and, above 24.6 eV, neutral helium will absorb He II Balmer continuum, two-photon continuum, and Ly$\alpha$ photons, as well as H I LyC. The function $f(\nu)d\nu$ in equation (23) represents the normalized probability per recombination that a photon is emitted in the interval $(\nu, \nu + d\nu)$. We stress that the recombination emissivity from intervening material is largely independent on the (poorly known) sizes and gas densities of the absorbers, and depends only weakly on the temperature of the emitting media.

*d) Recombination Continuum*     The rate of direct recombinations to the $n^2L$ level can be computed using the Milne relation (Osterbrock 1989). This, together with equation (23) for hydrogen-like ions yields

$$f(\nu)\alpha_{n^2L} = \frac{4\pi}{c^2}\left(\frac{h^2}{2\pi m_e kT}\right)^{3/2} \frac{2n^3}{Z^2}\nu^2 \sigma_{\rm HI}(\nu/\nu_{th}) e^{-h(\nu-\nu_{th})/kT}, \quad (27)$$

where all symbols have their usual meaning or have been defined in the text. Direct captures to the $2^2L$ level of He II result in a radiation continuum with specific volume emissivity equal to $2(N_{\rm HeIII}/N_{\rm HII}) \sim 2(N_{\rm He}/N_{\rm H}) \sim 1/6$ of the emissivity arising from free-bound captures to the ground state of H I .

*e) Ly$\alpha$ Line Radiation*     Inserting equation (23) into equation (2), and taking $f(\nu) = \delta(\nu - \nu_\alpha)$, where $\nu_\alpha$ is the Ly$\alpha$ frequency and $\delta(x)$ is the usual delta function, the specific intensity associated with Ly$\alpha$ line radiation can be written as ($\nu \leq \nu_\alpha$)

$$\begin{aligned}J_\alpha(\nu_o, z_o) &= \frac{h}{4\pi}\int_{z_o}^\infty dz\,\nu\delta(\nu-\nu_\alpha)\frac{(1+z_o)^3}{(1+z)^3}B(z,\nu_\alpha)e^{-\tau_{eff}(\nu_o,z_o,z)} \\ &= \frac{h}{4\pi}\left(\frac{\nu_o}{\nu_\alpha}\right)^2 B(z_{em},\nu_\alpha)(1+z_o)e^{-\tau_{eff}(\nu_o,z_o,z_{em})},\end{aligned} \quad (28)$$

where $z_{em}$ denotes the redshift of the Ly$\alpha$ line emitters, $1 + z_{em} \equiv (1+z_o)\nu_\alpha/\nu_o$. The effective recombination coefficient to the $2^2P$ level has been assumed to be independent of column density, since resonant absorption of Ly$\beta$ photons populates the $3^2P$ level, which subsequently decays via H$\alpha$ and two-photon emission. The population of the $2^2P$ level arising from downward cascading of $n > 3$ levels (following the resonant absorption of higher order Lyman lines) can be neglected (see Pengelly 1964 and Appendix A).

*f) Two-Photon Continuum*     The $2^2S$ level is populated by direct recombinations and by downward cascades following recombinations to higher levels. The function $g_{\rm HI}(\nu) = h\nu f(\nu)$ for the two-photon continuum decay of neutral hydrogen is tabulated in Osterbrock (1989). The emissivity due to the two-photon continuum decay of the $2^2S$ level of He II can be computed using the relation

$$g_{\rm HeII}(\nu) = g_{\rm HI}(\nu/4). \quad (29)$$

The effective recombination coefficient to the $2^2S$ level strongly depends on the Ly$\beta$ line center opacity, $\tau(\text{Ly}\beta)$. As this increases from 0 to $\infty$, $\alpha_{2^2S}^{eff}$ increases by a factor $\approx 2.6$. However, for clouds with $\tau \gtrsim 1$, which are the major contributors to the recombination emissivity, $\tau(\text{Ly}\beta) \gtrsim 10^3$. We have therefore assumed that the population of the $2^2S$ is well described by Case B recombination for all values of $N_{\text{HI}}$ (see Appendix A for details).

## 5. NUMERICAL RESULTS

The formalism we have developed so far, including the effects of the reprocessing of UV photons by a clumpy intergalactic medium, can be applied to compute the metagalactic flux arising from any discrete source of photoionizing radiation, the most plausible candidates being AGNs and young, star-forming galaxies. The last contribution is the most difficult to assess and also the most intriguing, as a determination of the UV background intensity and/or spectrum at high redshift might constrain the history of cosmic metal production and galaxy formation (Bechtold et al. 1987; Miralda-Escudé & Ostriker 1990; Songaila, Cowie, & Lilly 1990; Madau & Shull 1995). The intensity and spectrum of the ultraviolet background arising from massive stars in metal-producing galaxies will be the subject of a companion paper (Madau & Haardt 1995). Below, we will focus our attention on quasar sources.

*a) Quasar Emissivity as a Function of Redshift*   According to recent surveys (see, e.g., the one by Boyle, Shanks, & Peterson 1988, with about 400 ultraviolet-excess quasars), a pure luminosity evolution model, in which QSOs statistically conserve their number since $z \sim 2$, describes well the global properties of the quasar population at low redshift. Beyond $z \sim 2$, the comoving space density of quasars appears to stay constant up to a redshift of $\sim 3$ (see, e.g., Koo & Kron 1988; Boyle, Jones, & Shanks 1991). The decline in the quasar counts at $z \gtrsim 3$ has been the subject of a long-standing debate, and it is still controversial. It has been argued by Boyle (1991) and Irwin et al. (1991) that the space density of bright ($M_B < -26$) quasars at $z \sim 4.3$ is similar to that at $z \sim 2$. On the other hand, Warren, Hewett, & Osmer (1994) find strong evidence for positive evolution of the quasar luminosity function (LF) beyond $z \sim 2$. According to these authors, the evolution ceases at $z \sim 3.3$, with a marked decline by a factor $> 3$ in space density at larger redshift, as first suggested by Osmer (1982). A decline in the LF is also seen at the low luminosities ($M_B \gtrsim -26$) probed by the Schmidt et al. (1991) grism survey. We remark that these estimates are quite uncertain because of corrections for incompleteness and the adopted continuum emission spectrum, and because of the possible bias introduced by dust obscuration arising from intervening systems (Ostriker & Heisler 1984; Wright 1990; Fall & Pei 1993).

We shall estimate the integrated volume emissivity from QSOs following Pei (1995), who has recently derived an analytical model which fits well the empirical LF estimated by Hartwick & Shade (1990) and Warren et al. (1994). [2] The best-fit to the QSO LF is the standard double power-law:

$$\phi(L,z) = \frac{\phi_*}{L_*(z)} \left\{ \left[\frac{L}{L_*(z)}\right]^{\beta_1} + \left[\frac{L}{L_*(z)}\right]^{\beta_2} \right\}^{-1}. \tag{30}$$

---

[2] Hartwick & Shade (1990) have compiled 15 major optical surveys with more than 1000 quasars covering moderate redshifts $0.1 \lesssim z \lesssim 3.3$. Most recently, Warren et al. (1994) have combined their own and four other surveys to form a sample with contains about 200 quasars covering the range of high redshift $2 \lesssim z \lesssim 4.5$.

The entire LF shifts along the luminosity axis as the position of the break $L_*$ evolves with redshift:

$$L_*(z) = L_*(0)(1+z)^{\alpha-1} \exp[-z(z-2z_*)/2\sigma_*^2], \tag{31}$$

where a power-law spectral distribution for the typical quasar UV spectrum has been assumed, $f(\nu) \propto \nu^{-\alpha}$, and the dependence on the spectral index $\alpha$ is explicitly shown. According to Pei (1995), this Gaussian form can fit reasonably well the observational data in the entire range $0 \lesssim z \lesssim 4.5$; the evolution of QSOs reaches a maximum at $z \sim 2.8$ and declines at higher redshifts. In a cosmology with $h_{50} = 1$ and $q_0 = 0.1$, the LF fitting parameters values are $\beta_1 = 1.83$, $\beta_2 = 3.7$, $z_* = 2.77$, $\sigma_* = 0.91$, and $\log(\phi_*/\text{Gpc}^{-3}) = 2.37$; the B-magnitude at the break is $M_{B,*} = -23$. At the present epoch this LF is in good agreement with estimates of the Seyfert LF by Cheng et al. (1985). The proper volume emissivity can be written as

$$\epsilon_Q(\nu, z) = \epsilon_Q(\nu_B, 0)(1+z)^{\alpha+2}(\nu/\nu_B)^{-\alpha} \exp[-z(z-2z_*)/2\sigma_*^2], \tag{32}$$

where $\epsilon_Q(\nu_B, 0)$ is the extrapolated $z = 0$ emissivity at the reference frequency $\nu_B = c/4400$ Å:

$$\epsilon_Q(\nu_B, 0) = \int_{L_{min}}^{\infty} \phi(L, 0) L\, dL \simeq 6.4 \times 10^{32}\ \text{ergs Gpc}^{-3}\ \text{s}^{-1}\ \text{Hz}^{-1}. \tag{33}$$

The LF of Seyfert galaxies matches remarkably well that of optically selected QSOs at $M_B = -23$, and does not show clear evidence of leveling off down to $M_B \approx -18.5$ (Cheng et al. 1985), corresponding to $L_{min} \approx 1.5 \times 10^{-2} L_*(0)$. For the given shape of the LF, a value of $L_{min}$ 4 times smaller would increase the emissivity by only 20%.

*b) Quasar Spectral Energy Distribution* One of the biggest uncertainties in these calculations has always been the detailed shape of the UV continuum spectrum of quasars. Recently, however, the results of several spectroscopic surveys have significantly advanced our knowledge of the QSO emission properties. In particular, a very high signal-to-noise composite spectrum of the rest-frame ultraviolet and optical region of about 700 high luminosity, intermediate redshift quasars (part of the Large Bright Quasar Survey) has been constructed by Francis et al. (1991). Force-fitting a power-law to the constituent spectra yields a median spectral index of $\alpha_o \sim 0.3$ in the range 5000–1500 Å. In the UV, longward of the Ly$\alpha$ emission line, the survey of Sargent, Steidel, & Boksenberg (1989) yields a softer slope, $\alpha_{UV} = 0.78 \pm 0.27$ (see also O'Brien, Gondhalekar, & Wilson 1988). Finally, the Tytler et al. (1995b) *Hubble Space Telescope* FOC snapshot survey (82 QSOs at $z \sim 3$) has provided, for the first time, statistical information on the intrinsic ionizing photon distribution of quasar. After accounting for intervening absorption, Tytler et al. find an even softer slope, $\alpha_L = 1.4 \pm 0.25$, in the wavelength range 1300–330 Å.

We shall therefore adopt the following model for the "typical" quasar spectral energy distribution:

$$f(\nu) \propto \begin{cases} \nu^{-0.3} & (2500 < \lambda < 4400\ \text{Å}); \\ \nu^{-0.8} & (1216 < \lambda < 2500\ \text{Å}); \\ \nu^{-1.5} & (\lambda < 1216\ \text{Å}), \end{cases} \tag{34}$$

where the different slopes have been continuosly matched. In the far-UV, this model is similar to the "medium" QSO spectrum used in Bechtold et al. (1987), and "Model QS2" of Miralda-Escudé

& Ostriker (1990); its extrapolation to 2 keV gives $\alpha_{ox} \simeq 1.4$, in agreement with the value for QSOs measured by the *Einstein Observatory* (Zamorani et al. 1981). The soft (0.2–2 keV) X-ray spectral index is consistent with *ROSAT* observations of QSOs (e.g., Laor et al. 1994), but is significantly steeper than suggested by earlier *Einstein* and *EXOSAT* data (Wilkes & Elvis 1987; Comastri et al. 1992). Note that we have ignored luminosity and redshift dependent effects in QSO UV/soft X-ray emission spectra (O'Brien et al. 1988; Bechtold et al. 1994), and that the adopted X-ray spectral slope is too steep to fit the diffuse X-ray background below 40 keV (see, e.g., Madau, Ghisellini, & Fabian 1994 and references therein). Also, no attempt has been made to include the soft X-ray excess emission below 0.2 keV which has been detected in many AGNs (Turner & Pounds 1989) and is well modeled by the tail of a thermal spectrum with $kT \sim 40$–$80$ eV (Comastri et al. 1992). It has been suggested (see Sciama 1994) that the presence of this thermal, far-UV bump might help to account for the large abundances of N v and O vi observed in some Lyman-limit systems.

In Figure 4 we plot the (proper) volume emissivity at the H i , He i , and He ii Lyman edges due to QSOs as a function of redshift. Note that the "$K$-correction" term in equation (32) has been matched to take into account the varying spectral index $\alpha$ of equation (34). In the redshift range $0 \lesssim z \lesssim 1$, $\epsilon(912\,\text{Å},z)$ matches to within 30% the volume emissivity adopted in Madau (1992) from the Boyle (1991) quasar LF. At $z \approx 1.5$ (3) the former is about 2 (2.5) larger than the latter, then falls like a gaussian beyond $z \sim 4$ in contrast to the constant comoving emissivity scenario (Boyle 1991; Madau 1992).

*c) Spectrum of the Ionizing Background*   The formal solution of the radiative transfer equation (2) is not an explicit form of the specific background intensity, as its right-handed term implicitly contains $J$ in the emissivity term, $\epsilon = \epsilon_Q + \epsilon_r(J)$, and in the effective optical depth $\tau_{eff}(J)$. Physically, this simply means that the metagalactic flux depends on the ionization state of intervening clouds which is in turn determined by the background field itself. In order to solve for $J$, an iterative scheme must be implemented (see Appendix C). Figure 5 shows the background radiation field produced by the observed QSOs in a $q_0 = 0.1$ cosmology, as a function of redshift from the present epoch to $z = 5.5$. The spectrum from 5 to 5000 Å results from the numerical integration of equation (2), and is self-consistently filtered through a clumpy IGM. For comparison we have also plotted the same spectrum in the limit of a perfectly transparent ($\tau_{eff} = 0$) universe, and a purely absorbing intervening medium. As already discussed by a number of authors (Bechtold et al. 1987; Miralda-Escudé & Ostriker 1990, 1992; Madau 1991, 1992; Meiksin & Madau 1993; Zuo & Phinney 1993), the absorption by hydrogen and helium in intervening material produces a very strong effect in the spectrum of the ultraviolet metagalactic flux. This will be generally characterized by two breaks at the H i and He ii ionization edges, the amplitudes of which generally increase with redshift. The background intensity above 4 Ryd is, in particular, a non-linear function of the number of photons emitted by QSOs at these frequencies: a relative small decrease in this number causes a large increase in the size of the He ii break. The importance of the reprocessing of ionizing photons by intergalactic material is clearly seen in Figure 5. While redshifted He ii Ly$\alpha$ and two-photon continuum emission dominate the metagalactic flux from $\sim 1$ to 3 Ryd, direct recombinations to the ground level of hydrogen and singly ionized helium provide a substantial number of LyC ionizing photons close to the ionization thresholds. Because of the larger cosmic opacity at early epochs, the amplitude of the jump in the background intensity at 1216 Å and 304 Å due to H i and He ii Ly$\alpha$ emission increases with redshift from $z = 0$ to $z \approx 4$, to decrease again at $z > 4$ as the QSO emissivity drops, and the background flux becomes dominated by more local, less attenuated sources. At low redshift, Ly$\alpha$ line recombination radiation forms an extended hump in the background spectrum due to the integrated contribution of QSO absorbers along the line of sight. The diffuse Ly$\alpha$ flux becomes more peaked at early epochs, again because of the rapid decrease in the quasar ionizing emissivity above $z \sim 4$. Also note how, at $z \gtrsim 5$, helium starts to recombine: this produces a small absorption feature in the spectrum of the metagalactic flux at 504 Å.

*d) Photoionization Rates* The net effect of recombination emission from QSO absorbers on the ionization state of the universe is best depicted in Figure 6, where the total photoionization rates $\Gamma_{\rm HI}, \Gamma_{\rm HeI}$, and $\Gamma_{\rm HeII}$ are plotted as a function of redshift and compared to the case of a purely absorbing IGM.[3] Relative to the latter, our model calculations including riemission produce a value of $\Gamma_{\rm HI}$ about 1.5 times higher, nearly independent of redshift. Typically, less than 1/3 of the IGM diffuse contribution comes from optically thick Lyman-limit systems. The effect of recombination radiation makes $\Gamma_{\rm HeII}$ 1.6 times larger at $z = 1$, and 2.2 times larger at $z = 3$. How do these rates compare with observational constraints? Diffuse H$\alpha$ surface brightness observations of the Haynes-Giovanelli intergalactic cloud by Vogel et al. (1995) have recently yielded an upper limit to the local metagalactic photoionization rate of $3 \times 10^{-13}\,{\rm s}^{-1}$. The total QSO+IGM contribution at $z = 0$ from Figure 6 is about eight times lower, $\Gamma_{\rm HI} \simeq 4 \times 10^{-14}\,{\rm s}^{-1}$. Because of the larger QSO emissivity adopted here, and the inclusion of recombination radiation from intergalactic material, this value is about two times higher than the one obtained by Madau (1992). The amount of diffuse ionizing radiation predicted by our model at the present epoch is consistent with the value derived by Kulkarni & Fall (1993) by applying the proximity-effect technique towards QSOs at $\langle z \rangle \approx 0.5$. Observations at 21 cm of the fairly abrupt truncation in the H I distribution at the edge of nearby spiral galaxies have also been used to infer values of the metagalactic photoionzing flux. This technique is somewhat model-dependent, and yields rates in the range $10^{-14} - 3 \times 10^{-13}\,{\rm s}^{-1}$ (Maloney 1993; Corbelli & Salpeter 1993; Dove & Shull 1994), again consistent with our model calculations.

At high redshift, the proximity effect, which is the measured decrease in the number of Ly$\alpha$ -absorbing clouds induced by the UV radiation field of a QSO in its neighborhood, provides an estimate of the intensity of the metagalactic flux at the hydrogen Lyman edge. Determinations of $J_{912}$ in recent years include the following: $\log J_{912} \sim -20.5$ at $1.6 < z < 4.1$ (Bechtold 1994), between $-22$ and $-21.5$ at $z \approx 4.2$ (Williger et al. 1994), $\log J_{912} = -21 \pm 0.5$ at $1.7 < z < 3.8$ (Bajtlik, Duncan, & Ostriker 1988; Lu, Wolfe, & Turnshek 1991), and $\log J_{912} = -21.3 \pm 0.2$ at $1.7 < z < 3.8$ (for a velocity offset of $1400\,{\rm km\,s}^{-1}$ between the emission lines in the broad line region and the QSO rest-frame, see Espey 1993). If $J_\nu \propto \nu^{-1}$, the corresponding photoionization rate equals $3 \times 10^{-12} J_{912,-21}\,{\rm s}^{-1}$. Our QSO-dominated background model generates $\Gamma_{\rm HI} \approx 1.6 \times 10^{-12}\,{\rm s}^{-1}$ at $z = 2.5$, and a value three times lower at $z = 4$. *We conclude that, within the uncertainties, the observed QSOs appear able to provide the number of ionizing photons required by the proximity effect at $z \lesssim 4$.*

---

[3] A Gaussian fit, $\Gamma = A(1 + z)^B \exp[-(z - z_c)^2/S]$, yields for the H I photoionization rate plotted in Figure 6 $A = 6.7 \times 10^{-13}\,{\rm sec}^{-1}$, $B = 0.73$, $z_c = 2.30$, and $S = 1.90$. This fit can be applied over the range $0 < z < 5$ with a maximum error lower than 10%. A similar fitting formula to 10% accuracy in the same redshift range gives for the He II rate $A = 6.7 \times 10^{-15}\,{\rm sec}^{-1}$, $B = 0.72$, $z_c = 2.35$, and $S = 1.98$.

*e) Integrated* H I *Lyα Emission at z = 0 from Quasar Absorption Systems*     We now address the long-standing question of whether redshifted 1216 Å H I Lyα emission from intergalactic gas might be detected in the diffuse far-UV background in the range 1300–2000 Å. The intensity of the ionizing diffuse flux from quasar sources implies that the redshift-smeared background at $z = 0$ due to H I Lyα recombination radiation from photoionized Lyα clouds and Lyman-limit systems, about $3\,{\rm phot\,cm^{-2}\,s^{-1}\,Å^{-1}\,sr^{-1}}$ at 1500 Å, is at a level of less than five percent of current observational limits (see Bowyer 1991 for a review). Therefore, in a universe photoionized by QSOs, redshifted Lyα recombination emission from intergalactic material can be ruled out as a significant source of far-UV radiation.

*f) Intensity Fluctuations of the Ionizing Background*     The attenuation by QSO absorption-line systems reduces significantly the total number of primary sources involved in the production of the metagalactic flux. The qualitative behaviour can be readily obtained by expanding the effective optical depth in redshift around $z_o$. From equation (B9) we obtain

$$\Delta\tau_{eff}(912\,\text{Å}, z) \simeq \left[0.12(1+z)^{2.46} + 0.25(1+z)^{1.55}\right]\Delta z \qquad (35)$$

at the hydrogen Lyman edge, where the two terms on the right-hand side represent the contribution of the Lyα clouds and Lyman-limit systems to the effective opacity. At $z = 3$, $\Delta\tau_{eff}(912\,\text{Å}) \sim 1$ for $\Delta z \simeq 0.17$; in a low $q_0$ universe, this corresponds to a (proper) absorption length of $\Delta\ell \simeq 64 h_{50}^{-1}$ Mpc. The ionizing radiation field is thus largely local at 1 Ryd, as in the "attenuation volume" $4\pi(\Delta\ell)^3/3$ the QSO LF (eq. [30]) predicts only $\sim 4$ objects with $L > L_*(z = 3)$.

As argued by Zuo (1992) and Fardal & Shull (1993), this fact might have an interesting consequence, namely intergalactic absorption should greatly enhance the intensity and spectrum fluctuations of the ultraviolet background at high redshift. The importance of this effect resides in the possibility that the study of the influence of a fluctuating metagalactic flux upon Lyα absorption lines (e.g., the production of clearings and voids in the forest along the line of sight) could shed some light about the nature of the main contributors of ionizing background photons (Kovner & Rees 1989). Here we want to stress that, as recombination radiation from the numerous (relative to QSOs) Lyα clouds and Lyman-limit systems contributes $\sim 30\%$ of the total photoionization rate at $z \sim 3$, the UV background will be much smoother in space than previously considered. In particular, the two-point correlation function induced in the Lyα forest by intensity fluctuations will be reduced by a factor of about 2 (Fardal & Shull 1993).

*g)* H I *Gunn-Peterson Test*     In the presence of uniform material of density $n(z)$ along the path to a distant object, any resonance line with wavelength $\lambda$ will produce a Gunn-Peterson (1965) absorption trough with optical depth $\tau_{GP}(z)(= \pi e^2/m_e c) H(z)^{-1} \lambda f n(z)$, where $z$ is the redshift at which the observed radiation passes through the resonance, $H(z)$ is the Hubble constant, $f$ is the oscillator strength of the transition, and all other symbols have their usual meaning. The H I optical depth is related to the photoionization rate through the condition of ionization equilibrium

$$\tau_{GP}^{\rm HI} = 0.18 h_{50}^{-1} T_4^{-0.75} (\Omega_D h_{50}^2)^2 (1+z)^5 (1+2q_0 z)^{-1/2} \Gamma_{\rm HI,-12}^{-1}, \qquad (36)$$

where $T_4$ is the temperature of the ionized diffuse intercloud medium, $\Omega_D$ its mass density parameter, and $\Gamma_{\rm HI,-12}$ is the ionization rate in units of $10^{-12}$ s$^{-1}$. The strongest limit on the amount of diffuse intergalactic neutral hydrogen at $z \sim 3$ is provided by a direct estimate of the quasar Q2126–258 flux in regions of the spectrum where lines are absent (Giallongo, Cristiani, & Trevese 1992). The derived $1\sigma$ upper bound is $\tau_{GP}^{\rm HI} < 0.04$. Inserting this limit into equation (36) with $T_4 = 2$, $z = 3$, $q_0 = 0.1$, and $\Gamma_{\rm HI,-12} = 1.6$ yields $\Omega_D h_{50}^{3/2} < 0.027$. Nucleosynthesis constraints on the cosmological baryon density require $\Omega_B h_{50}^2 > 0.04$ (Walker et al. 1991), and would be consistent with the limit on $\Omega_D$ above if less than 70% of the baryons were in un uniformly distributed medium, the rest having collapsed into galaxies and discrete absorption systems. Two possible problems must be noted here:

1. If the diffuse IGM is photoionized by quasars sources only, the Gunn-Peterson test becomes more and more difficult to satisfy at earlier epochs, both because of the strong dependence of $\tau_{GP}$ with redshift and of the rapid drop in the QSO emissivity beyond $z \sim 4$. As depicted in Figure 6, the background ionization rate due to QSOs falls by more than a factor of 5 from $z \approx 3$ to $z \approx 4.3$, while the cosmological term in equation (36) increases by 4. Therefore we would expect the Gunn-Peterson optical depth to increase over 20 times. However, an analysis of the spectrum of the QSO 1202-0725 by Giallongo et al. (1994) yields $\tau_{GP} < 0.05$ at $z \simeq 4.3$.

2. In the gravitational instability scenario for the formation of cosmic structures, the assumption of a uniform intercloud medium has no theoretical justification. As recently argued by Reisenegger & Miralda-Escudé (1995), a photoionized IGM must be inhomogeneous on scales larger than the Jeans length, and the Gunn-Peterson opacity should fluctuate. Moreover, as the filling factor of underdense regions or voids is large, the median $\tau_{GP}$ should be significantly lower than the value expected for a uniform medium with $\Omega_D = \Omega_B$ (Reisenegger & Miralda-Escudé 1995).

*h) He II Gunn-Peterson Test* The recent *HST* observations of quasars Q0302-003 at $z = 3.29$ (Jakobsen et al. 1994) and PKS1935-692 at $z = 3.19$ (Tytler et al. 1995b), and the ASTRO-2 *HUT* spectrum of quasar HS1700+6416 at $z = 2.72$ (Davidsen et al. 1995), show the presence of absorption troughs at the redshifted wavelength of He II 304 Å. It is still unclear whether line blanketing from discrete Ly$\alpha$ forest clouds alone can produce the observed flux reduction, or whether a significant contribution from absorption in a smoothly distributed IGM is needed instead (see, e.g., Songaila, Hu, & Cowie 1995). In the latter case, the He II Gunn-Peterson test could potentially provide stringent constraints on the amplitude and spectral shape of the UV ionizing background (Meiksin & Madau 1993; Miralda-Escudé 1993; Madau & Meiksin 1994; Giroux et al. 1995). The He II and H I optical depths can be related as

$$\frac{\tau_{GP}^{\rm HeII}}{\tau_{GP}^{\rm HI}} = \frac{\eta_{thin}}{4}. \tag{37}$$

In Figure 7 we plot as a function of redshift the He II /H I ratio for an optically thin gas in ionization equilibrium with our model ultraviolet background. The predicted value increases from $\approx 25$ at $z = 0$ to $\sim 45$ at $z = 2.5$, to decrease again below 30 at $z \gtrsim 4.5$. This is because at low redshifts the universe is nearly transparent, while at $z \gtrsim 4.5$ the decline in the quasar counts makes the metagalactic flux dominated by more local, unattenuated sources. Note how recombination radiation from intergalactic material tends to decrease the expected value of $\eta_{thin}$ for $z > 1$.

A background dominated by quasar sources could then produce a He II Gunn-Peterson optical depth $\tau_{GP}^{\rm HeII} \sim 0.5$ at $z \sim 3$ and still be consistent with the lack (within the uncertainties) of any detectable $\tau_{GP}^{\rm HI}$. We conclude that the He II 304 Å absorption decrement observed by Jakobsen et al. (1994), Tytler et al. (1995b), and Davidsen et al. (1995) should be partly due to uniformly distributed He II , unless most of the baryons have been removed from the IGM, possibly into the Lyα forest.

*i) He II Lyα Clouds*   With the launch of *HUT* on ASTRO-2 it has become possible to detect and identify He II Lyα lines from Lyα forest clouds (Davidsen et al. 1995). These could provide, together with the corresponding H I Lyα lines, constraints on the spectrum of the metagalactic flux. A curve of growth analysis would enable us to estimate the parameter $\eta_{thin}$, and thus to test the QSO-dominated background scenario. Note that, because of the relatively soft diffuse radiation field produced by QSOs, He II lines associated with H I lines which are still in the linear part of the curve of growth are expected to be heavily saturated, even more so if massive stars in star-forming galaxies provide a significant contribution to the background flux.

*j) He II Lyα Line Blanketing*   At wavelengths shortward of the H I and He II Lyα lines, the quasar's continuum intensity is attenuated by the combined blanketing effect of many absorption lines arising from intervening discrete systems. As mentioned above, the relative strength of He II versus H I continuum and line opacity depends on the intensity of the metagalactic flux at 4 Ryd compared to 1 Ryd. The recent detection of He II Lyα absorption in the spectra of three high-$z$ quasars can thus be used as a diagnostic tool of the ambient physical conditions at early epochs. The effective H I line-blanketing optical depth due to Poisson-distributed absorbers is given by

$$\tau_{eff}^{\rm HI}(z) = \frac{1+z}{\lambda_\alpha} \int_{W_{min}}^{W_{max}} \frac{\partial^2 N}{\partial W \partial z} W dW \qquad (38)$$

(Paresce, McKee, & Bowyer 1980), where $\partial^2 N/\partial W \partial z$ is the observed rest equivalent width distribution. The He II effective optical depth may be obtained from equation (38) using a curve of growth analysis and a distribution of Doppler parameters $b$ for H I and He II . From the numerical study of Madau & Meiksin (1994), we derive for thermally broadened clouds ($b = {\rm const} \sim 30$ km) with $W > W_{min} = 0.0025$ Å ($N_{\rm HI} > 5 \times 10^{11}$ cm$^{-2}$)

$$\tau_{eff}^{\rm HeII}(z = 3.3) \approx 0.35 \eta_{thin}^{0.33}, \qquad (39)$$

a fit to 3% accuracy over the range $10 < \eta_{thin} < 100$. As our model background generates $\eta_{thin} \approx 45$ (corresponding to a "softness" parameter $S_L \equiv J_{912}/J_{228} \approx 0.6\eta_{thin} \sim 30$) at $z \approx 3.3$, we conclude that a He II Lyα line blanketing opacity greater than unity is to be expected in a universe photoionized by QSOs (Miralda-Escudé 1993; Madau & Meiksin 1994; Giroux et al. 1995). Note that, as the blanketing opacity is dominated by those lines which lie at the transition between the linear and the flat part of the curve of growth, $\tau_{eff}^{\rm HeII}$ is in fact sensitive to the undetermined H I distribution at very low H I equivalent widths, and increases if the hydrogen and helium absorption lines have the same width (turbulent broadening) instead (Madau & Meiksin 1994; Cowie et al. 1995).

*k) The Redshift Evolution of the Lyα Clouds*   As mentioned in § 2.2, the number of Lyα forest lines seen by the *HST* at $z \sim 0$ is larger than predicted by extrapolating from the high-$z$ regime (Bahcall et al. 1991; Morris et al. 1991). It has been suggested by Ikeuchi & Turner (1991) that this excess might be related to the sharp drop in the ionizing photon flux expected for $z < 2$ in a QSO-dominated UV background. We point out that a similar effect should be observable at $z \gtrsim 4$. As shown in Figure 6, the photoionization rate $\Gamma_{\rm HI}$ from QSOs is estimated to decrease by a factor of 6 between $z = 2.5$ and $z = 4.3$. If there is no compensating change in the evolution of cloud properties, this drop should produce an increase in the neutral hydrogen column and therefore in the number of Ly$\alpha$ clouds observed per unit redshift. A strong increase in the rate of incidence of Ly$\alpha$ clouds at large redshift has been claimed by Williger et al. (1994), who find $dN/dz \propto (1+z)^{4.6}$ in the range $3.7 < z < 4.2$.

*l) Photoionization Models for Metal-Line Absorption Systems*   It has been long recognized that it is possible to model the ionization structure of the heavy-element QSO absorption systems and discriminate between different background spectral shapes by looking at the line strengths of various elements (Chaffee et al. 1986; Bergeron & Stasinska 1986; Steidel & Sargent 1989). Measurements of the column densities of H I and several ionized species of C, N, and O in three optically thin systems by Vogel & Reimers (1993) have been used in conjuction with photoionization models to rule out star-forming galaxies as the main source of the UV background. The dominance of high-ionization species can be best explained by an AGN-type spectral energy distribution with softness $S_L < 5$. However, when combined with the presence of strong He I resonance lines (Reimers & Vogel 1993), these observations appear incompatible with photoionization equilibrium models (Giroux, Sutherland, & Shull 1994). Using Si II, Si IV, and C IV lines, Songaila et al. (1995) have recently estimated $S_L$ to be in the range between 20 and 100 at $z \approx 3.2$.

Here, we only want to point out the inadequacy of photoionization models which assume a background power-law ionizing spectrum. As clearly shown in Figure 5, the reprocessing of LyC photons by intergalactic material introduces bumps, wiggles, and breaks at characteristic wavelengths in the spectrum of the metagalactic flux, the presence of which must be taken into account when modeling metal systems. The integrated background due to He II Ly$\alpha$ line emission at 40.8 eV is clearly going to affect the abundance of, e.g., O II ions. We find that the O II /C III ratio is particularly sensitive to the spectrum of the metagalactic flux at 35 eV compared to 48 eV, and hence to He II Ly$\alpha$ recombination radiation (see Fig. 6). The absorption line O II 833 (together with C III 977) has been recently identified by Vogel & Reimers (1995) in the spectrum of the quasar HS1700+6416.

*m) A Case with $q_0 = 0.5$*   For $z \lesssim 1$, optical surveys are complete to a sufficiently faint magnitude level to detect the QSOs that contribute the bulk of the ionizing background, and the computed metagalactic flux does not depend on the assumed cosmology. However, at larger $z$ this will no longer be true, as more and more quasars fall below the survey limit. It is then necessary to correct for the missing QSOs: the selection function one adopts depends on the value of $q_0$. In the attempt to bracket the uncertainties on $J(\nu, z)$, we have run a $q_0 = 0.5$ background model using the quasar LF fitting parameters given in Pei (1995) for this cosmology. In a flat universe, the usual Gaussian fit, $\Gamma = A(1+z)^B \exp[-(z-z_c)^2/S]$ (good to 10% over the relevant redshift range) yields for the H I photoionization rate $A = 6.7 \times 10^{-13}\,{\rm sec}^{-1}$, $B = 0.43$, $z_c = 2.30$, and $S = 1.95$. Similarly, the parameters for the He II rate are $A = 6.3 \times 10^{-15}\,{\rm sec}^{-1}$, $B = 0.51$, $z_c = 2.3$, and $S = 2.35$. Relative to the $q_0 = 0.1$ case (see footnote [3]), the flat universe model provides about 30% less hydrogen- and helium-ionizing photons above $z = 2$.

*n) A Case with Low Intergalactic Absorption*   As already mentioned in § 2.2, a steepening in the $N_{\rm HI}$ power-law distribution of the Ly$\alpha$ clouds at high columns, suggested by several observations,

– 22 –may result in a smaller effective cosmic opacity and thus in an increased ionizing metagalactic flux (Meiksin & Madau 1993). In order to quantify this effect, we have run a $q_0 = 0.1$ background model using the same column density distribution of intervening absorbers given in equation (7), but assuming no forest clouds between $10^{16}$ and $1.59 \times 10^{17}\,\mathrm{cm}^{-2}$. The Gaussian fit to the new photoionization rates yields $A = 7.0 \times 10^{-13}\,\mathrm{sec}^{-1}$, $B = 0.8$, $z_c = 2.22$, and $S = 1.95$ for H I, and $A = 1.4 \times 10^{-14}\,\mathrm{sec}^{-1}$, $B = 0.69$, $z_c = 2.32$, and $S = 2.00$ for He II. Relative to the standard $q_0 = 0.1$ case (see footnote [3]), the low attenuation model provides about 30% (20%) more hydrogen-ionizing photons at $z = 0$ (2). However, the effect on the He II-ionizing flux is much larger, as the lack of Ly$\alpha$ forest clouds above $10^{16}\,\mathrm{cm}^{-2}$ significantly decreases the cosmic opacity above $54.4\,\mathrm{eV}$, thereby increasing $\Gamma_{\mathrm{HeII}}$ by about a factor of 2. Such a model would be characterized by a smaller He II over H I ratio, and may therefore have difficulties in accounting for a significant absorption trough at $304\,\mathrm{\AA}$ (Jakobsen et al. 1994).

## 6. SUMMARY

In this paper we have shown that, beside the attenuated direct radiation emitted from quasars and/or star-forming galaxies, the ionizing metagalactic flux will also have a significant diffuse component due to recombinations within the clumpy intergalactic gas. In a photoionized universe, the following processes will contribute to the diffuse hydrogen-ionizing background field from Ly$\alpha$ forest clouds and Lyman-limit systems:

1. Recombinations to the ground state of H I and He II.
2. He II Ly$\alpha$ emission ($2^2P \rightarrow 1^2S$) at $40.8\,\mathrm{eV}$.
3. He II two-photon ($2^2S \rightarrow 1^2S$) continuum emission.
4. He II Balmer continuum emission at $\geq 13.6\,\mathrm{eV}$.

In Case A, a fraction $\sim 0.7$ of the excited-state recombinations will populate the $2^2P$ state, while the remainder end up in the $2^2S$ state. In the presence of a power-law emissivity at a level currently envisaged from QSOs, the He II present in QSO absorption-line systems will efficiently degrade soft X-ray, He-ionizing photons into ultraviolet, H-ionizing photons. The addition of He I recombination radiation would be an improvement of this work, although this should only change our results at the very highest, $z \gtrsim 5$ redshift considered in this paper. We have modeled in detail the ionization state of intervening clouds and the propagation of ultraviolet radiation through intergalactic space, and shown that the universe is more opaque above 4 Ryd than previously estimated. We have also reassessed the contribution of the QSOs observed in optical surveys to the UV extragalactic background, and found that the attenuated direct quasar emission plus the recombination radiation from the intergalactic gas can provide enough hydrogen-ionizing photons to satisfy the proximity effect at large redshift. The reprocessing of LyC background photons by the intergalactic medium has been shown to significantly affect the intensity, spectrum, and fluctuations properties of the metagalactic flux.

The recent discovery by Cowie et al. (1995) and Tytler et al. (1995b) of metals in the Ly$\alpha$ clouds has shown that the IGM at high redshift is contaminated by the product of stars. In the context of our results, these measurements raise two important issues, which will be discussed in detail in a companion paper (Madau & Haardt 1995):

(1) As Ly$\alpha$ clouds have large filling factors and contain a significant fraction of the baryons in the universe, they might trace the bulk of star formation occurring in galaxies at high redshift.

Madau & Shull (1995) have shown that the background intensity of photons at 1 Ryd that accompanies the production of metals in the Ly$\alpha$ clouds may be significant, comparable with the estimated contribution from QSOs (unless a large fraction of the UV radiation emitted from stars cannot escape into the IGM, or if most of the metals observed at $z \approx 3$ were produced at much earlier epochs). If this is true, the signature of a stellar spectrum should appear in several places, notably the He II Gunn-Peterson effect and the distribution of metal ionization states of the Ly$\alpha$ forest and Lyman-limit absorbers.

(2) The new data could also imply the presence of a widespread distribution of dust in the IGM. Even a small amount of dust in the Ly$\alpha$ forest and Lyman-limit systems might provide a significant source of absorption of ionizing radiation in the universe. Further studies of this effect would be an improvement of the present work.

We thank A. Meiksin, J. Miralda-Escudé, J. Ostriker, Y. Pei, D. Tytler, and D. Weinberg for useful discussions, and N. Panagia for clarifying various issues concerning the physics of radiative transfer. This research was partially supported by the Director's Discretionary Research Fund at STScI. FH also acknowledges financial support by the Italian MURST and by the Swedish NFR Council.

# APPENDIX A

## RECOMBINATION RATE COEFFICIENTS

In this Appendix we provide some useful fitting formulae to the hydrogen recombination rate coefficients $\alpha_{n^2L}$ (in units of cm$^3$ s$^{-1}$) needed for our computations. The recombination coefficient to the ground level of hydrogen can be written as

$$\alpha_{1^2S} \simeq 1.58 \times 10^{-13} \times \begin{cases} T_4^{-0.51} & (T_4 \leq 1); \\ T_4^{-0.51-0.10 \ln T_4} & (T_4 \geq 1). \end{cases} \quad (A1)$$

This expression fits the results of Burgess (1964) within 3% for $0.1 \lesssim T_4 \lesssim 10$. We have also used the following fitting formulae (within 2% for $0.1 \lesssim T_4 \lesssim 10$) to the Case B results of Pengelly (1964):

$$\alpha_{2^2P}^{eff} \simeq 1.76 \times 10^{-13} \times \begin{cases} T_4^{-0.92-0.10 \ln T_4} & (T_4 \leq 1); \\ T_4^{-0.92-0.06 \ln T_4} & (T_4 \geq 1), \end{cases} \quad (A2)$$

and

$$\alpha_{2^2S}^{eff} \simeq 8.37 \times 10^{-14} \times \begin{cases} T_4^{-0.65-0.02 \ln T_4} & (T_4 \leq 1); \\ T_4^{-0.72-0.07 \ln T_4} & (T_4 \geq 1). \end{cases} \quad (A3)$$

# APPENDIX B

## RECOMBINATION EMISSIVITY

The integration over the H I column density distribution of QSO absorption-line systems appearing in the recombination emissivity term (eq. [24]) can be performed analitically over ranges of constant $\eta \equiv N_{\text{HeII}}/N_{\text{HI}}$ and for $\eta \propto 1/N_{\text{HI}}$ using approximated formulae for the absorption and escape probabilities. Taking

$$1 - e^{-\tau} \approx \frac{\tau}{1 + 0.75\tau} \quad (B1)$$

in the formulae for $p_{em}$, $w_{abs}$ and $\tilde{\alpha}$, the integral

$$K(\nu, \nu') = \int_{N_m}^{N_M} dN_{\text{HI}} \frac{\partial^2 N}{\partial N_{\text{HI}} \partial z} \frac{1}{\tilde{\alpha}(N_{\text{HI}})} p_{em}(N_{\text{HI}}, \nu) w_{abs}(N_{\text{HI}}, \nu') \quad (B2)$$

and the integral over the column density distribution which appears in the expression for the cosmic effective optical depth (eq. [5]) can be solved analytically. The computations are simple but lengthy. Though the right hand side of eq. (B1) does not have the right limit for large optical depths, it is a better approximation than $\tau/(1+\tau)$ for $\tau \lesssim 5$, with an error for $\tau \approx 1$ which is lower than 10%. The typical error of the analytical approximations presented in this Appendix is of order a few percent, with maximum values $< 15\%$.

*a) Effective Optical Depth*     The approximation given in eq. (B1) can be used to calculate the integral over the column density distribution which appears in eq. (5). When $\nu < 4\nu_L$, or when $\nu \geq 4\nu_L$ and the integration is over ranges of costant $\eta$ (i.e. for $N_{\rm HI} < N^*_{\rm HI} \equiv 1.2 \times 10^{18}$ cm$^{-2}$), we can write

$$\frac{d\tau_{eff}}{dz}(\nu) = \frac{8}{3} N_0 (1+z)^\gamma \frac{1}{\sqrt{r}} \arctan \sqrt{N_{\rm HI}/r} \ \Big|^{N_M}_{N_m}, \tag{B3}$$

where the constant $N_0$ and the exponent $\gamma$ are specified according to eq. (7). We have defined $r = 1/\Sigma(\nu)$, with $\Sigma(\nu) = [3/4\tau(\nu)]/N_{\rm HI}$. When $\nu \geq 4\nu_L$, and the integration is over ranges of $\eta \propto 1/N_{\rm HI}$ (i.e., for $N_{\rm HI} > N^*_{\rm HI}$), we have

$$\frac{d\tau_{eff}}{dz}(\nu) = \frac{8}{3} N_0(1+z)^\gamma \left\{ \left[\frac{1}{1+\Lambda(\nu)}\right] \frac{1}{\sqrt{r}} \arctan\sqrt{N_{\rm HI}/r} - \left[\frac{\Lambda(\nu)}{1+\Lambda(\nu)}\right] \frac{1}{\sqrt{N_{\rm HI}}} \right\} \Big|^{N_M}_{N_m}, \tag{B4}$$

where $\Lambda(\nu) = [3/4 N^*_{\rm HI} \eta_{max} \sigma_{\rm HeII}(\nu)]$. Here $r = [1 + \Lambda(\nu)]/\Sigma(\nu)$, with $\Sigma$ taking into account only the opacity due to H I and He I.

b) *H I Recombination Continuum* In computing the emissivity due to H I LyC radiation, we first note that the contribution of photons with $\nu' > 4\nu_L$ to the integral over $\nu'$ appearing in eq. (24) is negligible, and that the emissivity at $\nu > 4\nu_L$ is small compared to the emissivity due to He II LyC radiation at the same frequencies. Thus we can evaluate eq. (B2) for $\nu_L < \nu$ and $\nu' < 4\nu_L$, to obtain

$$K(\nu,\nu') = N_0(1+z)^\gamma \frac{2}{\alpha_B} r_2 r_3 \sigma_{\rm HI}(\nu') \sum_{i=1}^{3} \frac{R_i}{\sqrt{r_i}} \arctan\sqrt{N_{\rm HI}/r_i} \ \Big|^{N_M}_{N_m} \tag{B5}$$

for $\nu \neq \nu'$, where $\Sigma(\nu) = [3/4\tau(\nu)]/N_{\rm HI}$, $\alpha_B = \alpha_{\rm HeII} - \alpha_{1^2S}$, $r_0 = 1/\Sigma(1\,{\rm Ryd})$, $r_1 = \alpha_{\rm HeII}/[\Sigma(1\,{\rm Ryd})\alpha_B]$, $r_2 = 1/\Sigma(\nu')$, $r_3 = 1/\Sigma(\nu)$, and

$$\begin{aligned} R_1 &= [r_0 - r_1]/[(r_2 - r_1)(r_3 - r_1)]; \\ R_2 &= [r_0 - r_2]/[(r_1 - r_2)(r_3 - r_2)]; \\ R_3 &= [r_0 - r_3]/[(r_1 - r_3)(r_2 - r_3)]. \end{aligned} \tag{B6}$$

It is easy to obtain a similar formula when $\nu = \nu'$. In this case the second order multiplicity of the root $\Sigma(\nu)$ leads to a different decomposition of the integrand into integrable elementary partial fractions. We find

$$\begin{aligned} K(\nu',\nu') = &N_0(1+z)^\gamma \frac{2}{\alpha_B} r_2^2 \sigma_{\rm HI}(\nu') \times \\ &\left[ \sum_{i=1}^{2} \frac{R_i}{\sqrt{r_i}} \arctan\sqrt{N_{\rm HI}/r_i} + \frac{R_3 \sqrt{N_{\rm HI}}}{r_2 N_{\rm HI} + r_2^2} + \frac{R_3}{r_2 \sqrt{r_2}} \arctan\sqrt{N_{\rm HI}/r_2} \right] \Big|^{N_M}_{N_m}, \end{aligned} \tag{B7}$$

where now

$$\begin{aligned} R_1 &= [r_0 - r_1]/(r_2 - r_1)^2; \\ R_2 &= [r_1 - r_0]/(r_2 - r_1)^2; \\ R_3 &= [r_0 - r_2]/(r_1 - r_2). \end{aligned} \tag{B8}$$

*c) H I Lyα and Two-Photon Continuum*   The integral in eq. (B2) can be ulteriorly simplified when $p_{em}(\nu) = 1$, e.g., in the case of H I Lyα and two-photon emission. We obtain

$$K(\nu') = N_0(1+z)^\gamma \frac{2}{\alpha_B} r_2 \sigma_{\rm HI}(\nu') \sum_{i=1}^{2} \frac{R_i}{\sqrt{r_i}} \arctan \sqrt{N_{\rm HI}/r_i} \bigg|_{N_m}^{N_M}, \quad (B9)$$

where

$$\begin{aligned} R_1 &= [r_0 - r_1]/(r_2 - r_1); \\ R_2 &= [r_0 - r_2]/(r_1 - r_2). \end{aligned} \quad (B10)$$

The formulae (B5), (B7), and (B9) above are strictly valid for any integration range only when $h\nu' < 24.6$ eV. At higher photon energies, they are only correct as long as $N_M \leq N_{\rm HI}(4\,{\rm Ryd})$, as clouds with larger columns are characterized by a larger He I to H I ratio. The correct form of $w_{abs}(N_{\rm HI}, \nu')$ for these thicker clouds is

$$w_{abs}(\tau, \nu') = {\rm const} + [1 - {\rm e}^{-\tau(\nu')}] \frac{\tau_{\rm HI}(\nu')}{\tau(\nu')}. \quad (B11)$$

The integral over column in eq. (B2) can then be split into two distinct terms and solved.

*d) He II Recombination Continuum*   The evaluation of the emissivity due to He II recombination continuum can be simplified noting that the absorption of He II ionizing photons occurs at the He II photosphere. We also note that photons above 54.4 eV are effectively absorbed only by He II . The integral in eq. (B2) for $N_{\rm HI} > N_{\rm HI}(4\,{\rm Ryd})$ can then be written as

$$K(\nu, \nu') = \frac{1}{\alpha_A - 0.37\alpha_{1^2 S}} p_{em}(\nu) \frac{d\tau_{eff}}{dz}(\nu'), \quad (B12)$$

where $p_{em}$ is given by eq. (28). The result for $N_{\rm HI} < N_{\rm HI}(4\,{\rm Ryd})$ is analogous to eq. (B5), with $\sigma_{\rm HI}$ replaced by $\sigma_{\rm HeII}$ and $\Sigma$ evaluated at 4 Ryd in $r_0$ and $r_1$.

*e) He II Lyα and Two-Photon Continuum*   The arguments given above also imply that, in the case of He II Lyα and two-photon emission, $p_{em} \simeq 1$. The integral in eq. (B2) for $N_{\rm HI} > N_{\rm HI}(4\,{\rm Ryd})$ can then be written as

$$K(\nu') = \frac{1}{\alpha_A - 0.37\alpha_{1^2 S}} \frac{d\tau_{eff}}{dz}(\nu'). \quad (B13)$$

This becomes analogous to eq. (B9) for $N_{\rm HI} < N_{\rm HI}(4\,{\rm Ryd})$, again with $\sigma_{\rm HI}$ replaced by $\sigma_{\rm HeII}$ and $\Sigma$ evaluated at 4 Ryd in $r_0$ and $r_1$.

## APPENDIX C

### NUMERICAL INTEGRATION OF THE EQUATION OF RADIATIVE TRANSFER

The key problem to address for an iterative solution of the radiative transfer equation is how to achieve a fast numerical estimate of the effective optical depth $\tau_{eff}$ in each iteration. This is a three-dimensional function, since at any observed frequency $\nu_o$ it depends on the observer and emission redshifts $z_o$ and $z_{em}$. Computing time can be optimized noting that in the explicit form of $\tau_{eff}$ (eq. [5]), these three variables appear in the combination $\nu = \nu_o(1+z)/(1+z_o)$. A logarithmic redshift grid in the variable $x \equiv 1 + z$ is therefore designed, so that the ratio $R = x_{j+1}/x_j$ is constant for any $j < J$. We also set up a frequency grid longward of the H I and He II Lyman edges, defined by

$$\nu_i = \nu_L Z^2/R^{i-1} \qquad i = 1, 2, ...I. \tag{C1}$$

We make sure that the Ly$\alpha$ frequency coincides with one of the sampling points in this grid by properly adjusting the value of $R$, i.e., by setting $R = (4/3)^{1/m}$, where $m$ is an integer $\geq 1$. A third frequency grid provides coverage of the frequency range above 54.4 eV. The next step consists of computing the two sets of integrals which are needed to evaluate $\tau_{eff}$ as well as $d\tau_{eff}/dz$. The first of them,

$$U(i,j) = \int dN_{\rm HI} \frac{\partial^2 N}{\partial N_{\rm HI} \partial z}(1 - e^{-\tau_{ij}}), \tag{C2}$$

where

$$\tau_{ij} = N_{\rm HI}[\sigma_{\rm HI}(\nu_1 x_i/x_1) + \eta(x_j)\sigma_{\rm HeII}(\nu_1 x_i/x_1)], \tag{C3}$$

is computed for $1 \leq i \leq I$ and $j \geq i$. A second set of integrals,

$$L(i,j) = \int dN_{\rm HI} \frac{\partial^2 N}{\partial N_{\rm HI} \partial z}(1 - e^{-\tau_{ij}}), \tag{C4}$$

where

$$\tau_{ij} = N_{\rm HI}[\sigma_{\rm HI}(\nu_1 x_1/x_i) + \eta(x_j)\sigma_{\rm HeII}(\nu_1 x_1/x_i)], \tag{C5}$$

is evaluated for $1 \leq i \leq I$ and $1 \leq j \leq J$. Note that $U(1,j) = L(1,j)$, and $L(i,j) = 0$ longward of the H I Lyman limit. We can now calculate the effective optical depth $\tau_{eff}(\nu_o, z_o, z_{em}) \equiv T(i,j,k)$ ($j \leq k$) over the frequency$\times$redshift$\times$redshift space. Using the fact that the redshift distribution of the clouds can be expressed as a power-law (eq. [7]), we find the following recursive relation:

$$\begin{aligned} T(1,j,k) &= T(1,j,k-1) + [R^\gamma U(k-j+1,k) + U(k-j,k-1)]d_{k-1} & (k \geq 2); \\ T(i,j,j+1) &= [R^\gamma L(i-1,j+1) + L(i,j)]d_j & (i \geq 2); \\ T(i,j,k) &= T(i,j,j+1) + T(i-1,j+1,k), \end{aligned} \tag{C6}$$

where

$$d_i = x_i^{\gamma+1} \frac{(R-1)}{2}. \tag{C7}$$

Note that $T(i,j,j) = 0$, and $T(i,j,k) = 0$ for $k \leq i+j-1$ longward of the H I Lyman limit.

A computer code, named CUBA (from Cosmic Ultraviolet BAckground), has been designed based on the scheme sketched above. For each of the three frequency ranges, CUBA computes a number of integrals less that $2(I \times J)$. It is worth to point out that one should be careful in estimating the cloud emissivity (as described in Appendix B) at the photoionization threshold frequencies. Indeed, the opacity for photons at observed frequency $\nu_j < \nu_1$ is zero for $z_o \leq z \leq (\nu_1/\nu_j)z_o$, while the volume emissivity is non-zero for $z \geq (\nu_1/\nu_j)z_o$. The spectra shown in Figure 5 have been computed by CUBA over $35 \times 165$ time$\times$frequency grid points in a single run. Convergence (1 part in 1000) was reached after 16 iterations. Each iteration took about 15 seconds of CPU on a Digital VAX Alpha 3000/300.

# Figure Captions

**Figure 1:** *Solid curves:* The He II /H I ratio as a function of H I column density through a slab illuminated on both sides. The assumed total hydrogen gas density and temperature are $2\times 10^{-4}\,\mathrm{cm^{-3}}$ and $2.5\times 10^{4}\,\mathrm{K}$, respectively. The three curves refer, from top to bottom, to a metagalactic flux at the Lyman edge of intensity $5\times 10^{-22}$, $10^{-22}$, and $5\times 10^{-23}\,\mathrm{ergs\,cm^{-2}\,s^{-1}\,Hz^{-1}\,sr^{-1}}$. The spectral index is taken to be $\alpha = 2$. *Dashed curves:* Our "step-like" approximation (see text for details). *Dotted curves:* same ratio obtained by self-consistently solving the thermal structure of the clouds.

**Figure 2:** *Solid curves:* Absorption probability $(1-e^{-\tau})$ through an individual cloud as a function of neutral hydrogen column for four different ionizing photon energies (54.4, 75, 100 and 300 eV). *Dashed curves:* Same with our step-like approximation to the to the probability. *Dotted curves:* Same assuming a constant He II /H I value as derived in the optically thin limit. All curves are relative to the $J_{912} = 5\times 10^{-22}\,\mathrm{ergs\,cm^{-2}\,s^{-1}\,Hz^{-1}\,sr^{-1}}$ background model adopted for Figure 1.

**Figure 3:** *Upper panel:* Average temperature of a cloud as a function of neutral hydrogen column for the three background models adopted for Figure 1 (*solid line:* $5\times 10^{-22}$; *dotted line*: $10^{-22}$; and *dashed line*: $5\times 10^{-23}\,\mathrm{ergs\,cm^{-2}\,s^{-1}\,Hz^{-1}\,sr^{-1}}$). The gas total hydrogen density is taken to be $2\times 10^{-4}\,\mathrm{cm^{-3}}$ for $N_{\mathrm{HI}} < 1.59\times 10^{17}\,\mathrm{cm^{-2}}$, and $4\times 10^{-3}\,\mathrm{cm^{-3}}$ for thicker systems. *Lower panel:* Fractional population density of the $n=2$ H I level due to recombinations of photoionized ions (upper curves), collisional excitations of neutral atoms (middle curves) and recombinations of collisionally ionized ions (lower curves), as a function of the neutral hydrogen column.

**Figure 4:** The QSO proper volume emissivity at (from top to bottom) the H I , He I and He II ionization edges as a function of redshift. See text for details on the assumed QSO luminosity function and spectral energy distribution.

**Figure 5a:** The ionizing background from 5 to 5000 Å (*solid lines*) at epochs z=0, 0.5, 1 and 1.5, computed in a $q_0 = 0.1$ cosmology taking into account the absorption and reprocessing of QSO radiation by discrete absorption systems. The spectrum is compared to that arising from a perfectly transparent ($\tau_{eff} = 0$) universe (*dotted lines*), and a purely absorbing intervening medium (*dashed lines*).

**Figure 5b:** Same as Figure 5a but for z=2, 2.5, 3 and 3.5.

**Figure 5c:** Same as Figure 5a but for z=4, 4.5, 5 and 5.5.

**Figure 6:** *Upper panel:* Photoionization rates for H I (*solid line*), He I (*dashed line*) and He II (*dot-dashed line*) derived from our model background as a function of redshift. *Lower panel:* Ratios between these rates and the photoionization rates computed in the limit of a purely absorbing intervening medium. We also plot (again relative to the same quantity in the limit of a purely absorbing medium) the ratio between the O II and C III photoionization rates. As the O II and C III ionization potentials are 35 and 48 eV, respectively, this quantity will be particularly sensitive to He II Ly$\alpha$ line and two-photon continuum recombination radiation from intergalactic absorbers.

**Figure 7:** *Solid curve:* The He II /H I ratio as a function of redshift for an optically thin gas in ionization equilibrium with the ultraviolet background shown in Fig. 5. *Dashed curve:* Same in the limit of a purely absorbing medium.